\documentclass[11pt,a4paper]{article}
\usepackage{jcappub}

\newcommand{\p}{\partial}
\newcommand{\eq}{\begin{equation}}
\newcommand{\eqe}{\end{equation}}
\newcommand{\nn}{\nonumber}

\newcommand{\eqa}{\begin{eqnarray}}
\newcommand{\eqae}{\end{eqnarray}}

\title{Inflationary predictions in scalar-tensor DBI inflation}

\author[a,c]{Joel M. Weller,} 
\author[b]{Carsten van de Bruck}
\author[c]{and David F.~Mota}

\affiliation[a]{Institute for Theoretical Physics, University of Karlsruhe,\\
Karlsruhe Institute of Technology, 76128 Karlsruhe, Germany} 

\affiliation[b]{School of Mathematics and Statistics, University of Sheffield, \\
 Hounsfield Road, Sheffield S3 7RH, United Kingdom}
 
 \affiliation[c]{Institute of Theoretical Astrophysics, University of Oslo, \\
0315 Oslo, Norway}

\emailAdd{joel.weller@kit.edu}
\emailAdd{c.vandebruck@sheffield.ac.uk}
\emailAdd{d.f.mota@astro.uio.no}

\keywords{inflation, string theory and cosmology}

\abstract{
 The scalar-tensor Dirac-Born-Infeld (DBI) inflation scenario provides a simple mechanism to reduce the
large values of the boost factor associated with single field models with DBI action, whilst still being able to drive 60 efolds of inflation. 
Using a slow-roll approach, we obtain an analytical expression for the spectral index of the
perturbations and, moreover, determine numerically the regions of the parameter space of the model capable of giving rise to a power spectrum with amplitude and spectral index within the observed bounds. We find that regions that exhibit significant DBI effects throughout the inflationary period can be discarded by virtue of a blue-tilted spectral index, however, there are a number of viable cases --- associated with a more red-tilted spectral index ---  for which the boost factor is initially suppressed by the effect of the coupling between the fields, but increases later to moderate values. 

}
              
\begin{document}
\maketitle
\flushbottom

\section{Introduction}

Models of inflation based upon the Dirac-Born-Infeld (DBI) action \cite{Silverstein:2003hf,Alishahiha:2004eh} have been much 
studied in the past few years as an well-motivated example of
string cosmology.
In this context, the field $\chi$ appearing in the DBI Lagrangian is directly related to the
radial coordinate of a D3 brane moving in a `throat' region of a compactified space with a 
speed limit imposed upon its motion. The DBI model is an interesting example of k-inflation \cite{ArmendarizPicon:1999rj} (in which the
Lagrangian is an arbitrary function of the inflaton and its kinetic term) and has been 
much studied in this regard. K-inflationary models are noteworthy as the sound speed of the 
perturbations $c_s$ can be less than 1, which has the effect of forcing the
scalar perturbations generated during inflation to freeze-in at scales
shorter than the curvature radius \cite{Garriga:1999vw}. In the DBI model the sound speed is given in terms
of the boost factor $\gamma$ (which plays a role analogous to the Lorentz
factor in special relatively) by $c^2_s=\gamma^{-2}$, so $c_s\rightarrow 0 $
as the speed limit is saturated.
The original DBI papers concentrated mainly on this `relativistic' regime
where $\gamma$ is large and for which a simple power law inflation solution
can be obtained.
However, large values of the boost factor are inextricably linked with 
excessive amounts of non-Gaussianity in the bispectrum of the CMB
fluctuations as the non-linearity parameter $f_{\rm NL}$ is proportional to $\gamma^2$
in the relativistic case. 
This leads to an upper bound on the value of the boost factor $\gamma\lesssim 20$, assumed to apply during the first 10 of the last 60 efolds of inflation which directly affect the CMB radiation. 
 Since the boost factor increases with the number of efolds of inflation $N_{\rm max}$, many DBI models with $N_{\rm max}$ sufficiently large predict a high level of non-Gaussianity exceeding current observational bounds (cf. \cite{Bean:2007hc,Peiris:2007gz} and also \cite{Alishahiha:2004eh,Baumann:2006cd}). As shown in \cite{Lorenz:2008je} (cf. \cite{Lorenz:2008et}) including non-Gaussianity constraints
gives an upper bound on the slow-roll parameter. This greatly restricts the number of efolds of
inflation driven by the DBI field. 

One should also take into account the theoretical bound on the range of $\chi$, which arises simply because the throat must have finite size (sometimes referred to as the bulk-volume bound). 
This is related to the level of  tensor perturbations 
(given in terms of the the tensor-to-scalar ratio $r$) from inflation via the Lyth bound,
\begin{equation}
\Delta\chi = \int^{N_{\rm max}}_0 \left(\frac{r}{8}\right)^{1/2} dN
\end{equation}
which is true for both DBI and slow-roll inflation \cite{Baumann:2006cd,Lidsey:2007gq}.
This gives an upper bound on $r$ that can be combined with the lower bound derived from
the upper bound on $\gamma$ to place severe constraints on ultra-relativistic DBI models. 
Similar ideas are used in \cite{Spalinski:2007qy} to derive a series of consistency relations between the principle observable quantities.
The problems with observational constraints are made explicit in two detailed numerical studies \cite{Bean:2007hc} and \cite{Peiris:2007gz}, which use Monte Carlo methods to compare single field DBI models to recent cosmological observations. 
Both studies find that the majority of models cannot satisfy the bulk-volume bound; those that remain are slow-roll DBI models with small values of $\gamma$. 

Several routes have been explored that have a bearing on the problem, in particular the
inclusion of other fields in the determinant in the DBI Lagrangian to represent non-radial directions 
in the throat. 
With extra degrees of freedom, however, come extra features, in particular non-radial trajectories of the probe brane and non-adiabatic (entropic) fluctuations. These were addressed in \cite{Easson:2007dh} (see \cite{Gregory:2011cd} for recent developments), where the authors introduced `spinflation', accounting for angular momentum in the UV DBI model. Other models involving the angular coordinates of the DBI scenario have been considered in \cite{Langlois:2008wt, Langlois:2008qf} (see also \cite{Mizuno:2009cv,Mizuno:2009mv,Burgess:2005sb,RenauxPetel:2009sj,Langlois:2009ej}). A different type of multi-field model was introduced in \cite{Cai:2009hw} in which inflation is driven by two standard DBI fields, each corresponding to a brane with its own sound 
speed  (see also \cite{Chimento:2010un}, which generalises the assisted inflation scenario for multi-field DBI inflation). 
Another way in which the standard DBI scenario can be modified is by coupling the DBI field to another scalar field, or, as in \cite{Easson:2009kk,Easson:2009wc}, to the gravitational term in the action. 
The effect of coupling in multifield DBI models was considered in \cite{Brax:2009hd,Brax:2011si}, focusing on the effect of particle production due the interaction between the `inflaton' brane and trapped branes in the warped throat.\footnote{
 Although, like the scalar-tensor DBI inflation model in \cite{vandeBruck:2010yw}, this scenario involves a DBI action non-minimally coupled to a second field, the motivation and dynamics are very different: for example, in our model
 the energy density of the
 $\varphi$-field is not negligible (and indeed, is often the dominant contribution to the right-hand
 side of the Friedmann equation) and varies slowly as the DBI field evolves.
 These differences notwithstanding, something like this scenario might perhaps provide an promising
framework for the phenomenological model discussed here.
 }
 
In a previous work \cite{vandeBruck:2010yw}, we studied the consequences of coupling the DBI action to 
a canonical scalar field $\varphi$ in a scalar-tensor theory, in order to investigate the effect of an 
additional scalar field with a non-minimal coupling to gravity in the effective action.
We studied a coupling of the form $A=\exp(\beta\varphi)$, arising due to a conformal transformation to the Einstein frame.
As in the two-field model described in \cite{Brax:2008as}, 
when the coupling is non-zero, one field acquires a large effective mass, 
even for couplings of order a tenth of the 
strength of gravity ($\beta\gtrsim 0.1$).
The parameters
 of the DBI field, which are dependent on the additional scalar field, vary during inflation 
 so that the number of e-folds is extended and the boost factor is decreased (when compared to standard DBI inflation with the same bare parameters). 
 Two potentials for the canonical field were investigated, one with a minimum and one without,
 yielding similar results for both the background and perturbations,
 suggesting that the conclusions hold for any choice of potential steep enough to allow $\varphi$ to closely track the minimum of its effective potential.
 As discussed above, many DBI models predict a high level of non-Gaussianity arising due to large values of $\gamma$: in the coupled model, the boost factor is reduced, so the level of non-Gaussianity (which is proportional to $\gamma^2$  in single field DBI models) may be expected to be smaller than the standard DBI case.  If this is the case, the presence of the canonical scalar field may alleviate some of the problems of the DBI inflationary scenario,
 although we do not address the issue of non-Gaussianity in this paper.  

To understand the significance of the model, it is essential to calculate the predicted values
of any observable parameters so as to compare with data, however,
this is made rather difficult in this case as there are a large number of free parameters capable of affecting
the results.
In this article, we approach the problem from two fronts. After a brief resum\'{e} of the particulars
of the scalar-tensor DBI model in Sec. \ref{Sec:BGround}, we focus on the `slow-roll' limit of the model in 
Sec. \ref{AnalyticalPertTreatment}, showing that with a judicious choice of slow-roll parameters, the equations of motion for the perturbations can be made tractable and solved (to first-order) to obtain an expression for the
spectral index.
In Sec. \ref{Sec:DbiParameterSpace}, it is shown that the parameter space
can be severely constrained by considering the values of the power spectrum amplitude and
the spectral index. The latter measurement in particular rules out a large class of models ---
those with small values of $c_s$ at horizon crossing, which give
rise to a blue-tilted spectrum ---
so that only scenarios in which DBI characteristics are suppressed at the beginning of the 
observable period of inflation are allowed.
Finally, our results are summarised in Sec. \ref{Sec:DBI:Conclusions}

\section{Scalar-tensor DBI inflation}
\label{Sec:BGround}

In the Einstein frame (cf. \cite{vandeBruck:2010yw} for details) the scalar field $\varphi$ has a canonical kinetic term
and does not couple explicitly to the gravity sector, so the action for the model takes the form
\eq
S=\int d^4 x \sqrt{-g} \left[ \tfrac{1}{2}R + \mathcal{L}_\varphi + \mathcal{L}_{\rm DBI} \right] ,
\label{eq:Einsteinaction}
\eqe
where
\eqa
\mathcal{L}_\varphi &=& -\frac{1}{2}g^{\mu\nu}\varphi_{,\mu}\varphi_{,\mu}-U(\varphi), \\
 \mathcal{L}_{\rm DBI} &=& A^4(\varphi) \left[   f^{-1}(\chi)\left(1-\gamma^{-1}\right)-V(\chi) \right].
 \label{DBI_Lagrangian}
\eqae
$V(\chi)=\tfrac{1}{2}m^2\chi^2$ and $U(\varphi)$ are the potentials for the DBI field $\chi$ and the canonical scalar field $\varphi$
respectively. Two forms for $U(\varphi)$ were investigated in \cite{vandeBruck:2010yw},
 one of an
exponential form
\eq
U(\varphi) = U_0 \exp(-\eta\varphi), \label{ExpPot}
\eqe
and another, quadratic potential $U(\varphi) = U_0 (\varphi-\eta)^2$ with a minimum at $\varphi=\eta$. It was found that when the coupling between the fields was
non-zero, both potentials led to extremely similar behaviour, so for simplicity the exponential 
potential (\ref{ExpPot}) will be used throughout the body of the article, and the results of 
the numerical analysis for the quadratic potential will be summarised in Appendix \ref{QuadraticPotentialAppendix}.

$f(\chi)$ is the warp factor of the DBI field, a measure of the geometry of the throat region of
the compactified space in which the brane moves. As in \cite{vandeBruck:2010yw},
 in the following the `mass-gap solution' \cite{Alishahiha:2004eh,Kecskemeti:2006cg}
 \begin{equation} \label{muDef}
 f(\chi) = \frac{\lambda}{(\mu^2+\chi^2)^2},
 \end{equation}
 will be used, which exhibits 
the salient features of a capped throat with a cutoff at the IR end.
The presence of the mass scale $\mu$ (which must be subplanckian in order to get a warped throat) suppresses the DBI effects as the $\chi$-field moves to smaller values.  Generally, the smaller the value of $\mu$, the larger the $\gamma$ factor can become
during the inflationary period and the more efolds one observes.

The boost factor $\gamma$ appearing in (\ref{DBI_Lagrangian}) is modified by a $\varphi$-dependent factor of the conformal coupling $A(\varphi)$ compared
to the standard DBI model by a factor, so that 
\eq \label{eq:gam}
\gamma = \frac{1}{\sqrt{1-A^{-2}f\dot\chi^2}},
\eqe
where the dot indicates a derivative with respect to cosmic time. (Here, as throughout this
article, a flat Friedmann-Robertson-Walker metric is assumed: $ds^2 = -dt^2 + a^2(t)\delta_{ij}dx^i dx^j$.) The 
conformal coupling $A(\varphi)$ is given by
\eq \label{eq:Adef}
A(\varphi) = \exp(\beta\varphi),
\eqe
in terms of the constant $\beta$.

The dynamical evolution of the system is determined by the
 background equations of motion for the homogeneous fields 
 \eqa
 \ddot\chi &+& 3 H\gamma^{-2}\dot\chi + \tfrac{1}{2}A^2\frac{f_\chi}{f^2}(1-3\gamma^{-2}+2\gamma^{-3}) + A^2\gamma^{-3}V_{\chi}=
-\beta\dot\chi\dot{\varphi}(3\gamma^{-2}-1),
\label{eq:Echi} \\
\ddot{\varphi} &+& 3H\dot{\varphi}+U_{\varphi}=
\beta A^4 \left[ f^{-1}( 4-3\gamma^{-1}-\gamma )-4V \right],
\label{eq:Ephi}
 \eqae
 where $f_\chi = \p f/\p\chi$ and  $V_\chi = \p V/\p\chi$, together with the Friedmann equations
 \eqa
 3H^2 &=& \tfrac{1}{2}\dot{\varphi}^2+U+A^4\left[f^{-1}(\gamma-1)+V\right], 
  \label{eq:Fried1} \\
-2\dot{H} &=& \dot{\varphi}^2+\gamma A^2\dot\chi^2. \label{eq:Fried2}
 \eqae

\section{Analytical treatment of the perturbations}
\label{AnalyticalPertTreatment}

In \cite{vandeBruck:2010yw},
 the equations of motion for the first-order perturbations and the perturbed
Einstein equations were derived in full and solved numerically.
 As was expected, the perturbations of the DBI
field were found to propagate with a sound speed defined by $c^2_s=\gamma^{-2}$;
however, many extra terms appear
involving factors of $c_s$, the coupling $\beta$ and derivatives of the warp factor $f$, which 
render
the full equations intractable. In this section, we derive
approximations for the background dynamics in the scenario
in which the $\varphi$-field is trapped in the minimum of its effective potential.
It is shown that in this case one can rewrite the perturbation equations in
terms of slow-roll parameters and hence obtain an expression for the
spectral index of scalar perturbations that is in excellent agreement with the
numerical analysis in the following section.

\subsection{Defining slow-roll parameters}

As was shown in \cite{vandeBruck:2010yw}, in the coupled case with $\beta>0$, the situation
 is simplified considerably as the extra mass term on the right-hand side
 of (\ref{eq:Ephi})
 means that the effective potential can have a minimum value at which the
field value $\varphi=\varphi_{\rm min}$ satisfies,
\eq \label{eq:MinCond}
\frac{dU}{d\varphi}\bigg|_{\varphi=\varphi_{\rm min}}-\beta e^{4\beta\varphi_{min}} T_{DBI}^{b}=0,
\eqe 
where $T_{DBI}^{b} = f^{-1}(4-\gamma-3\gamma^{-1})-4V \approx -4V$ is the trace of the `bare' DBI stress-energy tensor and the $\varphi$ dependence of $A(\varphi)$ has been written explicitly using (\ref{eq:Adef}).
With the exponential potential (\ref{ExpPot})
the condition for the minimum in this case is 
\eq  \label{eq:ExpMinEqn}
U \simeq \frac{4\beta}{\eta}A^4V,
\eqe
so the minimum is given by the logarithmic function
\eq
 \varphi_{\rm min} = \frac{1}{4\beta+\eta}\log\left(\frac{\eta U_0}{4\beta V}\right),
\eqe
(cf. \cite{Brax:2008as}, in which a similar condition is obtained for two coupled scalar fields). 
Even for relatively small values of the coupling $\beta$, the $\varphi$-field is forced into
this minimum, the position of which changes as the $\chi$-field evolves as\footnote{
This corrects a minor error in eqn. (3.31) of \cite{vandeBruck:2010yw}.
}
\eq \label{FirstPhiApprox}
\dot\varphi = -\left( 1+\tfrac{4\beta}{\eta} \right)^{-1}\left( \frac{V_\chi}{V} \right)\frac{\dot\chi}{\eta},
\eqe
so that the second Friedmann equation can be approximated as
\eq
 -2\dot H \simeq A^2\gamma\dot\chi^2. 
\eqe
When the coupling between the two fields is switched on,
the DBI equation of motion is dominated by the Hubble
friction and potential terms (we have checked this numerically and found
it to be an excellent approximation). Thus (\ref{eq:Echi}) 
can be approximated by
the slow-roll equation
\eq \label{SlowRollChi}
3H\gamma^{-2}\dot\chi + A^2\gamma^{-3}V_\chi \simeq 0,
\eqe
i.e.
\eq \label{ChiApprox}
\dot\chi \simeq -\frac{A^2 V_\chi}{3H\gamma}.
\eqe
Finally, as the contribution of the kinetic terms of the fields in (\ref{eq:Fried1}) is negligible, one can rewrite the Friedmann equation as
\eq
3H^2 \simeq \left(1+\tfrac{4\beta}{\eta}\right)A^4 V, 
\eqe
using (\ref{eq:ExpMinEqn}).
We are now in a position to define the first slow roll parameter $\epsilon = -\dot H/H^2$, which
using (\ref{ChiApprox}),
can be written
\eq \label{EpsDef}
\epsilon \simeq \frac{1}{2A^2\gamma } \left(1+\tfrac{4\beta}{\eta}\right)^{-2} \left( \frac{V_\chi}{V} \right)^2.
\eqe
We can use this expression to obtain an expression for $\dot\varphi$ in terms of $\epsilon$.
Substituting (\ref{ChiApprox}) into (\ref{FirstPhiApprox}) gives
\eq \label{SecondPhiApprox}
\dot\varphi \simeq \tfrac{2}{\eta}H\epsilon.
\eqe
The rate of change of the slow-roll parameters should be
second order in $\epsilon$. Differentiating (\ref{EpsDef}) with respect to time
gives
\eqa
\frac{\dot\epsilon}{2H\epsilon} &\simeq& \frac{1}{2H\epsilon}\left[ \left(\frac{\dot c_s}{c_s} - 2\beta\dot\varphi \right)\epsilon
+\frac{1}{A^2\gamma}\left( 1+\tfrac{4\beta}{\eta} \right)^{-2} \left( \frac{V_\chi}{V} \right)
\left( \frac{V_{\chi\chi}}{V} - \frac{V_\chi^2}{V^2} \right)\dot\chi \right], \nn\\
&\simeq&   \frac{\dot c_s}{2H c_s} - \frac{2\beta}{\eta}\epsilon 
-\frac{1}{A^2\gamma}\left( 1+\tfrac{4\beta}{\eta} \right)^{-1}\left( \frac{V_{\chi\chi}}{V} - \frac{V_\chi^2}{V^2} \right), \nn\\
&\simeq&   \frac{\dot c_s}{2H c_s}
-\frac{1}{A^2\gamma}\left( 1+\tfrac{4\beta}{\eta} \right)^{-1}  \frac{V_{\chi\chi}}{V}
+2\left( 1+\tfrac{3\beta}{\eta} \right)\epsilon, \nn\\
&=& \tfrac{1}{2}s -\delta + 2\left( 1+\tfrac{3\beta}{\eta} \right)\epsilon,
\label{dotEps}
\eqae
where the slow-roll parameters $\delta$ and $s$ are defined as follows
\eq
\delta \equiv \frac{1}{A^2\gamma}\left( 1+\tfrac{4\beta}{\eta} \right)^{-1}  \frac{V_{\chi\chi}}{V},
\qquad s \equiv  \frac{\dot c_s}{H c_s}.
\eqe
$\delta$ is the equivalent of the 
commonly used $\eta$ parameter and $s$ is a measure of the
rate of change of the sound speed, which matches the definition used in the
DBI literature (cf. \cite{Baumann:2006cd}). Note, that
$s$ has an implicit dependence of $A(\varphi)$ through the definition of $\gamma$
in (\ref{eq:gam}).

\subsection{Rewriting the equations of motion}

It was demonstrated in \cite{vandeBruck:2010yw} that when the $\varphi$-field is trapped in the minimum of its effective
potential, the effect of the perturbations $\delta\varphi(t,{\bf x})$ is negligible and the coupled
DBI model is effectively a single field system. The first-order perturbations can then be well-described
by the Fourier components of the single variable $u = r_\chi Q_\chi $. Here $r_\chi$ is the combination of background quantities
\eq
r_\chi = aAc_s^{-3/2}
\eqe
and $Q_\chi \equiv \delta\chi+\frac{\dot\chi}{ H}\Psi$ is the gauge-invariant Mukhanov-Sasaki variable constructed from the DBI field perturbation $\delta\chi$ and the metric perturbation $\Psi$. The perturbation equation
in conformal time $\tau$ (with $' = d/d\tau$) is therefore [cf. eqn. (4.22) in \cite{vandeBruck:2010yw}]
\eq \label{First_uk_eqn}
u_k'' +  \left[ k^2c_s^2 + a^2C_{\chi\chi} - \frac{r_\chi''}{r_\chi} \right]u_k = 0,
\eqe 
with 
\eqa
C_{\chi\chi}&=& 
\frac{A^4\dot\chi}{ H}\frac{f_\chi}{f^2}(1-c_s)^2
-\left[ \frac{f_\chi}{f} -\frac{A^2\dot\chi}{ H c_s}\right]\frac{\dot{c_s}}{c_s}\dot\chi
-\tfrac{1}{2}c_s f_\chi A^{-4} \dot\chi^2 V_{T,\chi}  + 
\nn\\ %
&{}& 
+\tfrac{1}{2}A^2 (1-c_s)^2 \left[ c_s \left(\frac{f_\chi}{f^2}\right)_{,\chi} +(1+c_s)f^{-1}\left(\frac{f_\chi}{f}\right)_{,\chi} \right]
+\tfrac{3}{2}A^2\dot\chi^2 c_s^{-1}(1+c_s^2)-
\nn\\ %
&{}& 
-A^4 c_s^{-2}\frac{\dot\chi^4}{2 H^2}
-A^2 c_s^{-1}(1+c_s^2)\frac{\dot\chi^2\dot{\varphi}^2}{4 H^2}
+\frac{\dot\chi V_{T,\chi}}{ H}(1+c_s^2)
+c_s^{3}A^{-2} V_{T,\chi\chi}.
\label{CXX}
\eqae
Analysing each term in (\ref{CXX}) separately, one can rewrite the expression
 in terms of the slow-roll parameters $\epsilon$, $\delta$ and $s$ as
\eq \label{CChiChi}
C_{\chi\chi} \simeq 3H^2\left( \delta -s -2\epsilon\left[1+\tfrac{\beta}{\eta} \left( 1-c_s^2 \right) \right]\right).
\eqe
The interested reader is directed to Appendix \ref{DerivationAppendix} for full details of the derivation of this equation.
Differentiating $r_\chi$ with respect to conformal time gives
\eqa
\frac{r_\chi'}{r_\chi}  &=& \mathcal{H} \left(1+ \frac{\beta \varphi'}{\mathcal{H}} -\tfrac{3}{2} \frac{c_s'}{c_s \mathcal{H}} \right), \nn\\
&\simeq&   \mathcal{H} 
\left(1+ \tfrac{2\beta}{\eta}\epsilon-\tfrac{3}{2}s \right).
\eqae
Differentiating again (again assuming the derivatives of $\epsilon$ and $s$ are second order) yields
\eqa
\frac{r_\chi''}{r_\chi}  &=& \left(\frac{r_\chi'}{r_\chi} \right)' +  \left(\frac{r_\chi'}{r_\chi} \right)^2,  \nn\\
&\simeq&
 \mathcal{H}'\left(1+ \tfrac{2\beta}{\eta}\epsilon-\tfrac{3}{2}s \right)
+ \mathcal{H}^2\left(1+ \tfrac{4\beta}{\eta}\epsilon-3s \right), \nn\\
&\simeq&
 \mathcal{H}^2\left[ (1-\epsilon)\left(1+ \tfrac{2\beta}{\eta}\epsilon-\tfrac{3}{2}s \right)
 +\left(1+ \tfrac{4\beta}{\eta}\epsilon-3s \right) \right], \nn\\
&\simeq&
 \mathcal{H}^2\left[ 2 - \tfrac{9}{2}s -\left(1-\tfrac{6\beta}{\eta} \right)\epsilon \right],
 \label{RChiChiEq}
\eqae
using $\mathcal{H}' = \mathcal{H}^2(1-\epsilon)$. 
Combining (\ref{CChiChi}) and (\ref{RChiChiEq}) means the combination that appears in the equation of motion can 
be simplified to
\eq
 a^2C_{\chi\chi} - \frac{r_\chi''}{r_\chi} \simeq
 -\mathcal{H}^2\left[ 2 - \tfrac{3}{2}s -3\delta +\left(  5 + \tfrac{6\beta}{\eta}\left( 2-c_s^2 \right) \right)\epsilon \right].  \label{ApproxWithH2} 
\eqe
Defining the background variable $z$ by
\eq \label{zeq}
z=\frac{a\sqrt{p+\rho}}{c_sH} = \frac{a\sqrt{-2\dot H}}{c_sH} = a\gamma\sqrt{2\epsilon},
\eqe
one can show [using (\ref{dotEps}) and (\ref{ApproxWithH2})] that 
\eq \label{z''zeq}
a^2C_{\chi\chi} - \frac{r_\chi''}{r_\chi} \simeq -\frac{z''}{z} +\tfrac{6\beta}{\eta}\epsilon\mathcal{H}^2(1+c_s^2),
\eqe
which is correct to first order in $\epsilon$, $s$ and $\delta$. The first term on the RHS appears (without the factor of $A$ in the definition of $z$)
in single field k-inflation models \cite{Garriga:1999vw}. Alone, this would indicate
that the combination $|u_k / z |$ becomes approximately constant for small $k$; the small correction proportional to $(1+c_s^2)$ arises due to the fact that (\ref{First_uk_eqn}) is an
approximation to the full, two-field model given by eqns (4.21) and (4.22) in 
 \cite{vandeBruck:2010yw}. 
The correction is proportional to $\epsilon$ and thus subdominant to $z''/z$ in the mass term: this can be seen when comparing the size of the terms in the full equations (when evaluated numerically, as in the following section) and in many cases the correction is not too much larger than the terms proportional to the scalar field perturbations that have been neglected in (\ref{First_uk_eqn}).
Thus, to a good approximation, we can neglect the second term in (\ref{z''zeq}).
In this case the evolution of $u_k$ is determined by the (approximate) equation
\eq \label{Approx_uk_eqn}
u_k'' +  \left[ k^2c_s^2 -\frac{z''}{z}  \right]u_k \simeq 0,
\eqe
or, in terms of the slow-roll parameters
\eq  \label{Approx_uk_eqn2}
u_k'' +  \left( k^2c_s^2 -\mathcal{H}^2\left[ 2 - \tfrac{3}{2}s -3\delta +\left(  5 + \tfrac{18\beta}{\eta} \right)\epsilon \right]   \right)u_k \simeq 0.
\eqe
The comoving curvature perturbation of the full two-field system is given by eqn. (4.27)
 in \cite{vandeBruck:2010yw} as
 \[
 \mathcal{R} = \frac{H}{(-2\dot H)}\left[ \dot\varphi Q_\varphi + A^2\gamma\dot\chi Q_\chi \right].
 \]
where $Q_\chi$ and $Q_\varphi$ are the gauge-invariant Mukhanov-Sasaki variables for 
the perturbations of the $\chi$ and $\varphi$ fields respectively. Neglecting the $Q_\varphi$ term
and using (\ref{zeq}), one gets
\[
\mathcal{R} \simeq \frac{A\gamma^{1/2}}{\sqrt{2\epsilon}}Q_\chi = \frac{u_k}{z},
\]
as $ Q_\chi = u_k/r_\chi = \sqrt{2\epsilon}c_s^{1/2}A^{-1} u_k/z $.
The power spectrum is given by
\eq \label{PSpecExpr}
\mathcal{P}_\mathcal{R} \simeq \frac{k^3}{2\pi^2}\left| \frac{u_k}{z}\right|^2.
\eqe

\subsection{Solving the perturbation equation}

As in the standard slow-roll case, eqn. (\ref{Approx_uk_eqn2}) admits a solution in terms of Hankel functions. However, to account for the
fact that $c_s$ is not constant we follow the 
approach of \cite{Tzirakis:2008qy} and 
rewrite the equation using the time variable
\eq
y\equiv\frac{c_sk}{aH}=\frac{c_sk}{\mathcal{H}},
\eqe
so $y=1$ at sound horizon crossing. For a related approach, see \cite{Lorenz:2008et,Ringeval:2009jd}.
The derivatives of $u_k$ can be rewritten in terms of the slow-roll parameters as
\eq
\frac{du_k}{d\tau} = -c_s k (1-\epsilon-s) \frac{du_k}{dy}
\eqe
(using $\mathcal{H}'=\mathcal{H}^2(1-\epsilon)$) and
\eq
\frac{d^2u_k}{d\tau^2} = \mathcal{H}^2\left[ (1-\epsilon-s)^2 y^2 \frac{d^2u_k}{dy^2}-s(1-\epsilon-s)y\frac{du_k}{dy} \right].
\eqe
Substituting these expressions into (\ref{ApproxWithH2})
gives
\eq
(1-\epsilon-s)^2 y^2 \frac{d^2u_k}{dy^2}-s(1-\epsilon-s)y\frac{du_k}{dy}
+\left[ y^2 -2 + \tfrac{3}{2}s +3\delta -\epsilon\left(  5 + \tfrac{18\beta}{\eta} \right) \right]u_k=0,
\eqe
which can be written in the form
\eq \label{NewBessel}
y^2 \frac{d^2u_k}{dy^2}+(1-2p)y\frac{du_k}{dy}
+\left[\ell^2y^2+p^2-\nu^2 \right]u_k=0 
\eqe
with
\eqa
p &=& \tfrac{1}{2}(1+s), \\
\ell &=& (1-\epsilon-s)^{-1}, \\
\label{correct_nu}
 \nu &=& \tfrac{3}{2}+s-\delta+3\epsilon\left( 1+\tfrac{2\beta}{\eta} \right).
\eqae 
The solution of eqn. (\ref{NewBessel}) is $u_k = z^p\zeta_\nu(\ell z)$ where
$\zeta_\nu$ is a Bessel function of order $\nu$ \cite{abramowitz+stegun}.
The general solution is 
\eq
u_k(y) = y^{\tfrac{1}{2}+\tfrac{s}{2}}\left[ c_1 H^{(1)}_\nu(\ell y) + c_2 H^{(2)}_\nu (\ell y) \right]
\eqe
using Hankel functions of the first and second kind.
Comparing with the short wavelength solution $u_k(y\gg1)\approx e^{-kc_s\tau}/\sqrt{2c_s k}$
means
$c_2=0$ and $c_1 = \sqrt{\pi/4kc_s}$ so (up to a phase factor)
\eq
u_k(y) = \frac{1}{2}\sqrt{\frac{\pi}{c_sk}}\sqrt{\frac{y}{1-\epsilon-s}} H^{(1)}_\nu\left(\frac{y}{1-\epsilon-s} \right).
\eqe
In the long wavelength limit $H^{(1)}_\nu(y\ll 1) \sim \sqrt{\frac{2}{\pi}}e^{-i\pi/2}2^{\nu-\tfrac{3}{2}}\frac{\Gamma(v)}{\Gamma(3/2)}y^{-\nu}$,
so
\eq \label{eqforu}
|u_k(y)| \sim 2^{\nu-\tfrac{3}{2}}\frac{\Gamma(v)}{\Gamma(\tfrac{3}{2})}(1-\epsilon-s)^{\nu-\tfrac{1}{2}}
\frac{y^{\tfrac{1}{2}-\nu}}{\sqrt{2c_sk}}.
\eqe
 \begin{figure}[t!] 
\centering
\includegraphics[width=0.6\textwidth]{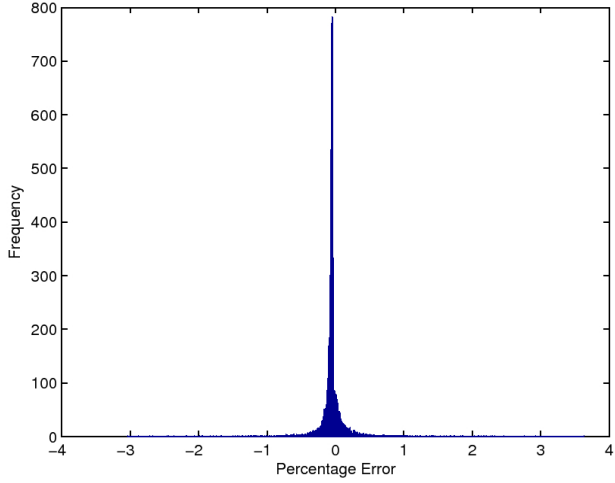}
\caption{For comparison between the numerical and analytical predictions,
the value of $n_s$ was estimated using (\protect\ref{Correct_ns}) for 13056 parameter sets
with $\mathcal{P}_\mathcal{R}$ in the observed range. The histogram shows the percentage error between the estimated result and that obtained from numerical integration of the
perturbation equations. It can be seen that the
vast majority of the $n_s$-values estimated using (\protect\ref{Correct_ns}) differ less than 1\% from the
calculated value.}
\label{fig:nsErrHist}
\end{figure}
%
%
Thus, from (\ref{PSpecExpr}) we find
\eq
\mathcal{P}_\mathcal{R}^{1/2} \simeq \frac{\sqrt{2c_sk}Hy}{2^{3/2}\pi\sqrt{c_s\epsilon}}|u_k|.
\eqe
Defining $\mathcal{V(\nu)} = 2^{\nu-3}(1-\epsilon-s)^{\nu-\tfrac{1}{2}}
\Gamma(v)/\Gamma(\tfrac{3}{2})$
and using (\ref{eqforu})
this can be written
\eq
\mathcal{P}_\mathcal{R}^{1/2} \simeq \left( \frac{\mathcal{V}(\nu)}{\pi} \right)
\frac{H}{\sqrt{c_s \epsilon}}y^{\tfrac{3}{2}-\nu},
\eqe
so the spectral index is 
\eqa
n_s &=& 1+(3-2\nu), \nn\\
&=& 1-2s+2\delta-6\epsilon\left( 1+\tfrac{2\beta}{\eta} \right). \label{Correct_ns}
\eqae
This estimate compares extremely well (significantly less than 1\% error in most cases) with the value obtained by solving the
perturbation equations numerically described in the following section. The comparison is
illustrated in Fig. \ref{fig:nsErrHist}. 
Note that, as $\beta/\eta$ is not necessarily less than one, the terms in $\epsilon\beta/\eta$ are
of the same order as the other slow-roll parameters and should not be neglected in (\ref{Correct_ns}). 
In fact, numerically, we find that dropping these terms shifts the value of the spectral index by a large factor of roughly 5-10\%.
Comparing (\ref{Correct_ns}) with the corresponding equation in single field slow-roll $n_s=1+2\delta-6\epsilon$ \cite{LiddleLyth} and the single field DBI case\footnote{
 This result is quoted [cf. eqn. (40) in \cite{Baumann:2006cd}] in the form $n_s=-2\epsilon-\tilde\eta-s$ where $\tilde\eta = \dot\epsilon/(H\epsilon)$. However, one can see from (\ref{dotEps}) that in the limit $\beta\rightarrow 0$, 
$\tilde\eta = s-2\delta+4\epsilon$, which gives the stated result.
 } 
 $n_s=1-2s +2\delta-6\epsilon$ \cite{Baumann:2006cd}, it can be seen that the standard expression
 is recovered when the coupling is small.

 In Fig. \ref{fig:nsErrHist}, (\ref{Correct_ns}) was evaluated at sound horizon crossing $kc_s = aH$ i.e $y=1$. 
 As the definitions of $\epsilon$ and $\delta$
 explicitly include factors of $A=e^{\beta\varphi}$, it is useful to check to what extent the power spectrum and the spectral index  depends on the evolution of the fields. 
 The sound horizon crossing formalism can be expressed by the 
condition
\eq \label{ExactSoundHorizon}
\frac{d}{dy}\left(\frac{H}{\sqrt{c_s \epsilon}}y^{\tfrac{3}{2}-\nu}  \right)=0,
\eqe
\cite{Tzirakis:2008qy} which ensures that the 
 that the expression for $\mathcal{P}_\mathcal{R}^{1/2}$
 is independent of $y$ and can safely be evaluated at the
 sound horizon crossing $y=1$.  Following \cite{Tzirakis:2008qy}, we can generalise this
 condition. 
 Observing that
 \eq
 \frac{d}{dy}=-\frac{1}{yH(1-\epsilon-s)}\frac{d}{dt},
 \eqe
 it follows that
 \eqa
 y\frac{d}{dy}\left(\frac{H}{\sqrt{c_s \epsilon}}y^{\tfrac{3}{2}-\nu}  \right) &=
& \left[ (\tfrac{3}{2}-\nu)\frac{H}{\sqrt{c_s \epsilon}}
 -\frac{1}{H(1-\epsilon-s)}\frac{d}{dt}\left( \frac{H}{\sqrt{c_s \epsilon}} \right) \right] y^{\tfrac{3}{2}-\nu}, \nn\\
 &\simeq& 
 \left[ \tfrac{3}{2}-\nu+\frac{\epsilon+\tfrac{1}{2}s+\dot\epsilon/(2\epsilon H)}{1-\epsilon-s} \right] \frac{H}{\sqrt{c_s \epsilon}}y^{\tfrac{3}{2}-\nu}, \nn\\
 \Rightarrow  \frac{d}{d\log (y)}\log \left(\frac{H}{\sqrt{c_s \epsilon}}y^{\tfrac{3}{2}-\nu}  \right) 
  &\simeq& 
  \tfrac{3}{2}-\nu+\frac{s-\delta+3\epsilon\left( 1+\tfrac{2\beta}{\eta} \right)}{1-\epsilon-s}.
 \eqae
Inserting the value of $\nu$ in (\ref{correct_nu}), obtained by neglecting the two-field correction to the $u_k$ mass term, it can be seen that (to first order in the
 slow-roll parameters) the expression for the power spectrum is independent of the $y$-value at which it is evaluated. 
This justifies our assumption that we can 
evaluate the power spectrum at sound horizon crossing. 

\section{The parameter space of the model}
\label{Sec:DbiParameterSpace}

The scalar-tensor DBI inflation model has thus far been investigated in detail in
 scenarios in which the $\varphi$-field is in the minimum of its 
effective potential, in order to specify uniquely the initial conditions and to neglect short-term effects due to the field oscillating as it reaches the minimum\footnote{
In this case the initial value of the $\chi$-field is of less importance 
as, for a given set of parameters, the system proceeds along a single trajectory in field space. The anisotropies in the CMB radiation are affected by the range of scales (around 4 orders of magnitude or 10 efolds) that cross the curvature horizon roughly 55 efolds before the end of inflation. Thus, only the behaviour of the system in the last 60-70 efolds of inflation is needed to compare with observations and we are free to choose values $\chi_{\rm ini}$ that allow an inflationary period of sufficient duration,
bearing in mind string theory constraints on the maximum length of the throat (cf. \cite{Baumann:2006cd,Bean:2007hc}) that mean $\chi_{\rm ini}\lesssim \mathcal{O}(1)$ in Planck units. In all of the examples presented here, $\chi_{\rm ini} = 1.5$ has been used.
}.
Even in this restricted case, the inflationary dynamics are determined by six parameters: the coupling $\beta$, the t'Hooft coupling $\lambda$ and the cutoff scale $\mu$ in the DBI warp factor, the mass term in the quadratic potential of the DBI field $m$ and the two constants in the $\varphi$-field potential $U_0$ and $\eta$.
It is possible to understand the importance of these different parameters by keeping five fixed and varying 
one (as in figs. 5, 7 and 9 in \cite{vandeBruck:2010yw}). 
However, this only gives a limited view of the effect of changing a particular parameter on the observations.
To improve upon this, one can perform multiple runs using random sets of parameters from within a given range to compare the resulting values of the power spectrum amplitude and spectral index with observations. 
In the following, the results of such an analysis are analysed for the exponential potential.
The quadratic potential yielded similar results: these are discussed in Appendix \ref{QuadraticPotentialAppendix}. 

\begin{figure}[t] 
	\begin{center}
	 	\begin{tabular}{cc}
			   \includegraphics[width=0.5\textwidth]{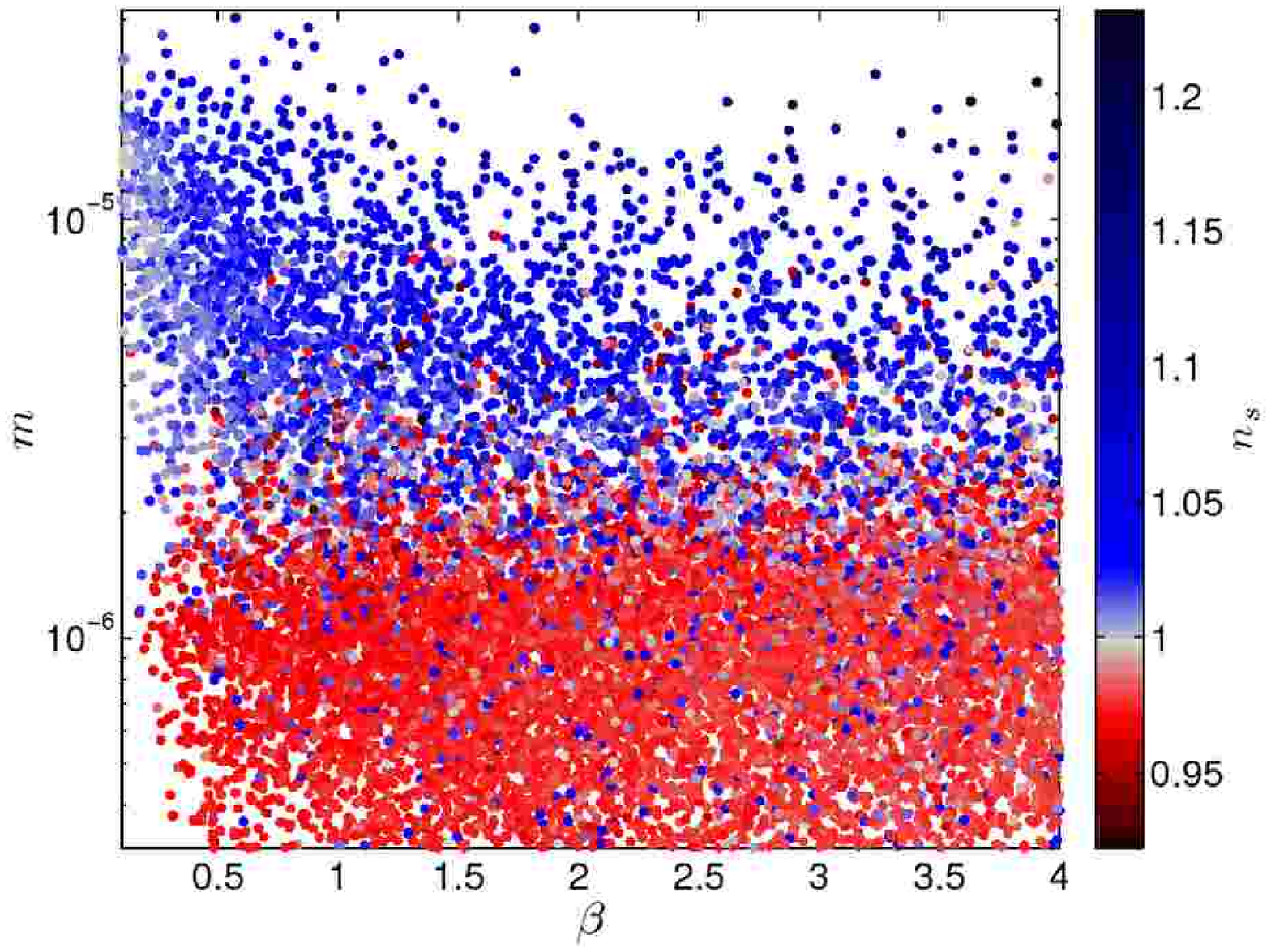} 
			& \includegraphics[width=0.5\textwidth]{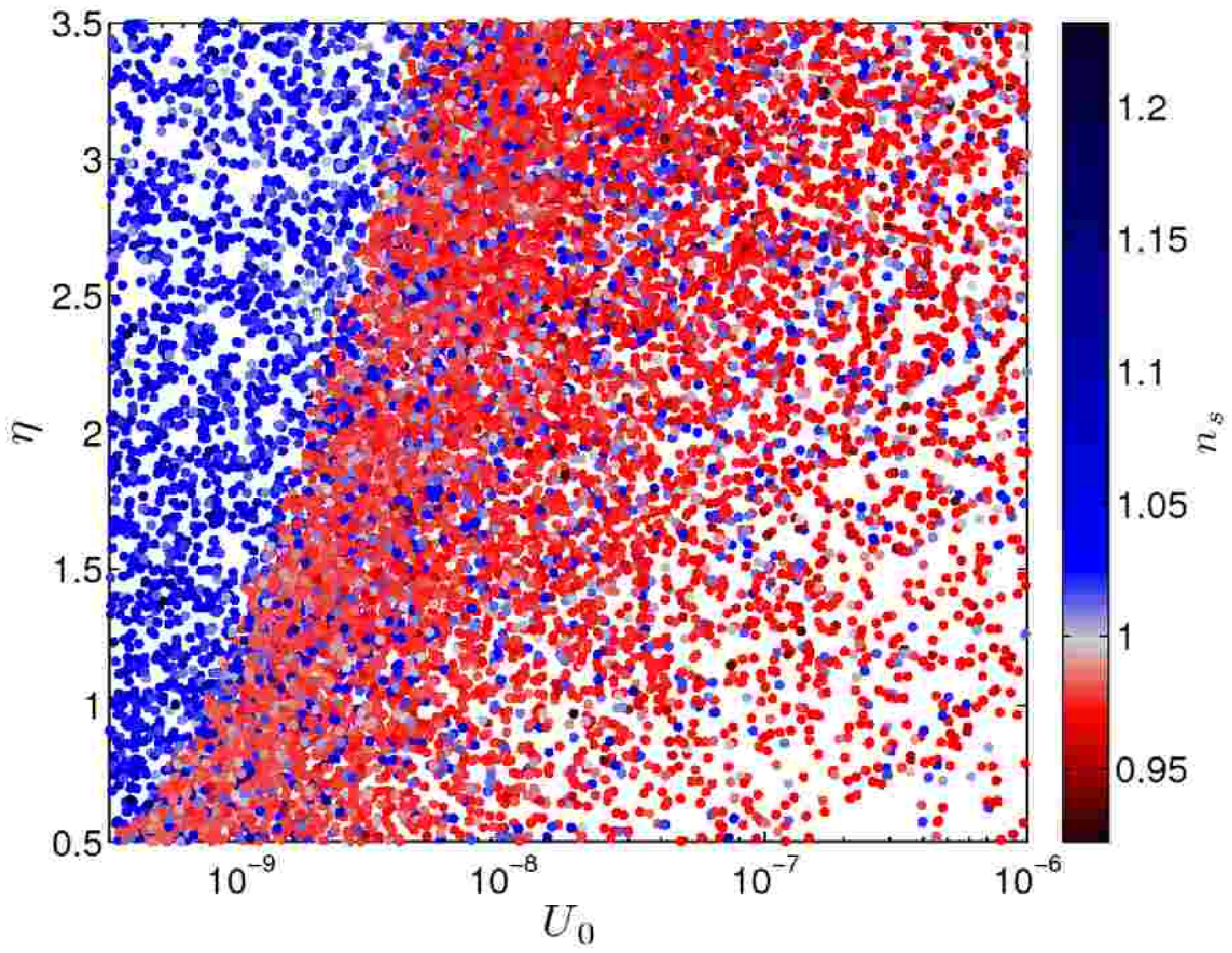} \\
			    \includegraphics[width=0.5\textwidth]{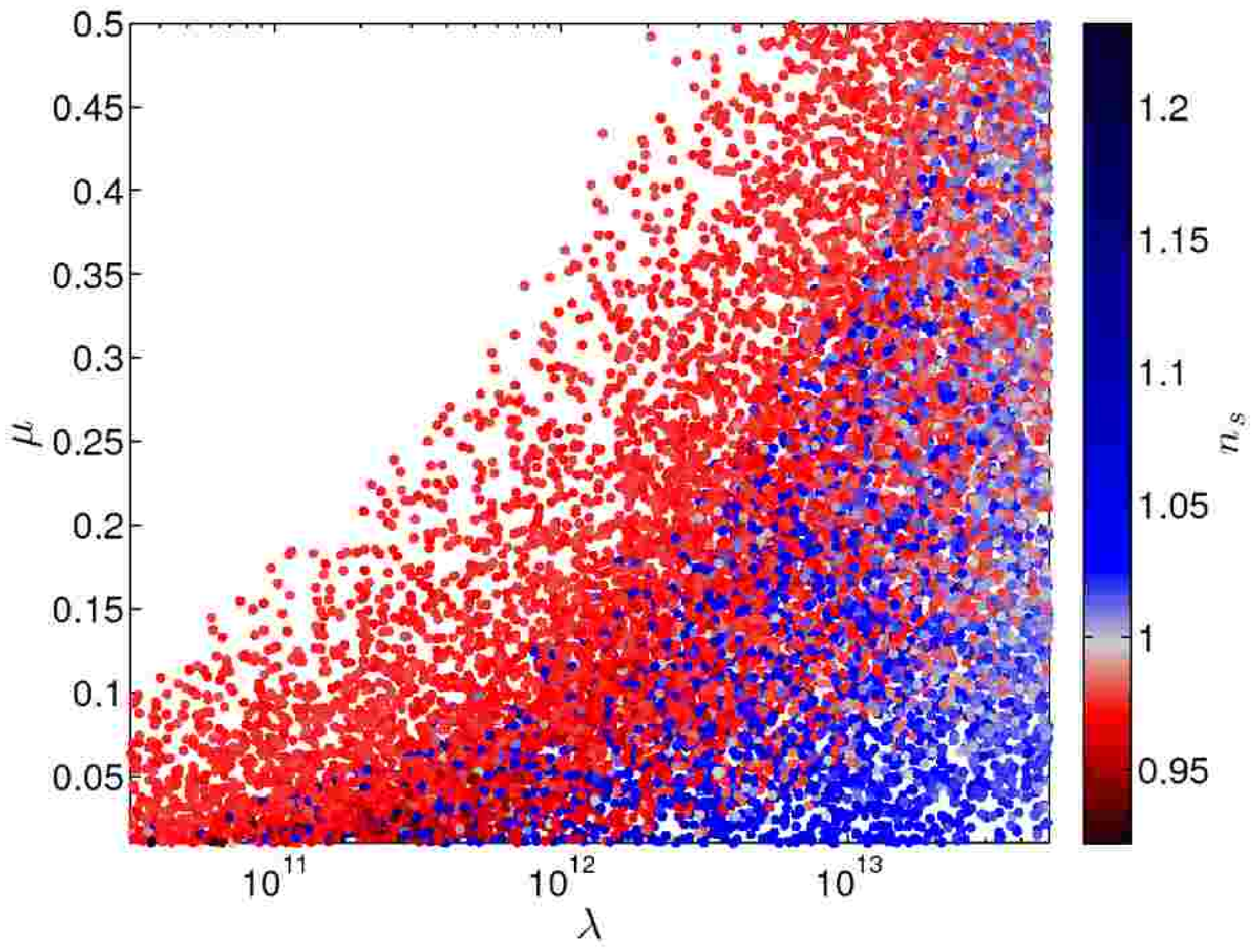}
			& \includegraphics[width=0.5\textwidth]{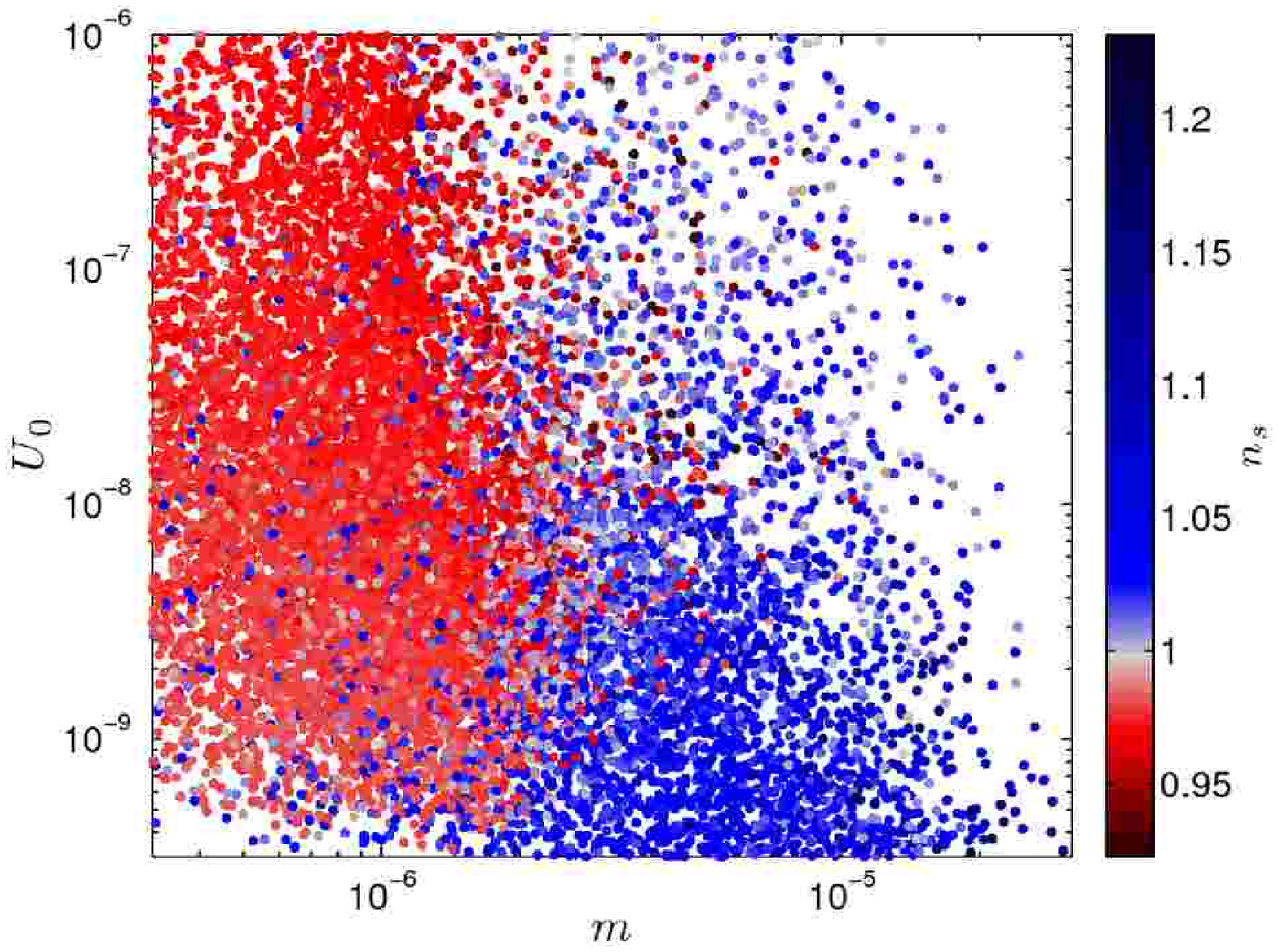}
		\end{tabular}
	\end{center}
	\caption[Spectral index plots for the `better' set 
	(with power spectrum amplitude within observational limits) 
	for the exponential potential.]
	{
	Spectral index plots for the `better' set (with power spectrum amplitude within observational limits) 
	for the exponential potential, colour-coded according to the value of $n_s$.
	Upper-left panel: the $\beta-\log_{10}m$ plane; upper-right panel: 
	the $\log_{10}U_0-\eta$ plane;
	lower-left panel: $\log_{10}\lambda-\mu$ plane; lower-right panel: $\log_{10}m-\log_{10}U_0$ plane.
	}
	\label{Exp_Better_ns}
\end{figure}

\begin{figure}
	\begin{center}
		\begin{tabular}{c}
		\includegraphics[width=1\textwidth]{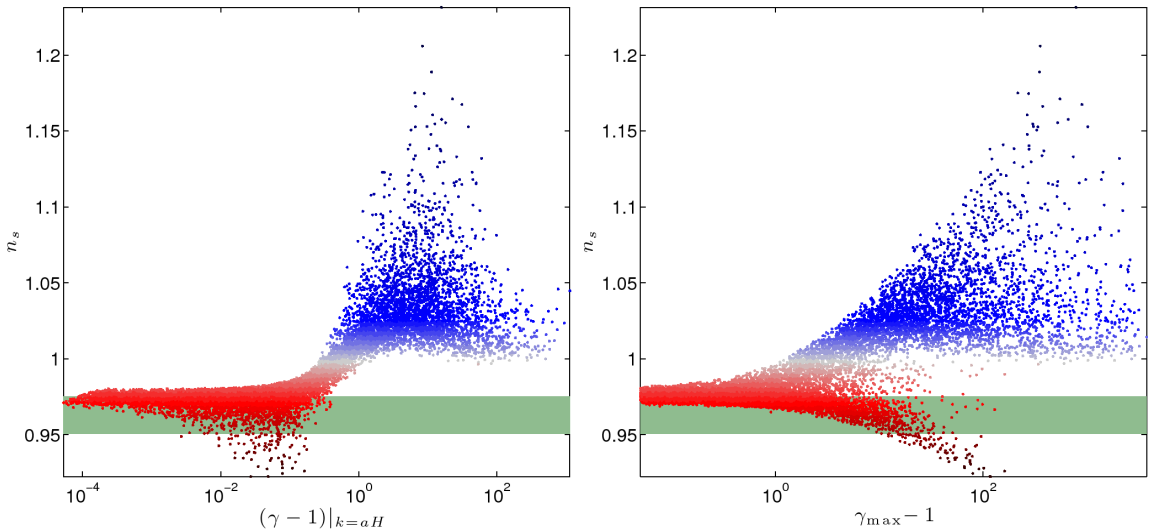} \\
		\end{tabular}
	\end{center}
		\caption{
		Correlations between the spectral index and the 
		DBI boost factor for the exponential potential (Note the log scale on the x-axis.)
		The left panels show $(\gamma-1)$ 
		(calculated at $k_{\rm pivot}=aH$) against $n_s$, with the
		green region indicating the observational limits on the spectral index.
		The right-panels show the correlation between $(\gamma_{\rm max}-1)$ and $n_s$. 
		(Points are 	
		colour-coded according to the value 
		of $n_s$, for easy comparison with the parameter-space
		plots in figs. \protect\ref{Exp_Better_ns}).
	 }
	 \label{Gam_ns}
\end{figure}

The parameter ranges\footnote{
These ranges were based on preliminary tests in which it was found that a large range of values of the
parameters $U_0$ and $\lambda$ (and to some extent, $m$) gave rise to a power spectrum amplitude within
observable limits. Only non-zero couplings $\beta>0$ were considered so as to ensure
the presence of a minimum for the $\varphi$-field, in which case there is a unique
trajectory in field space for a given set of parameters. 
}
used in the run with the exponential potential were
\eqa
\beta \in [0.1 ,\; 4],  & \lambda \in [10^{10.5} ,\;  5\times10^{13}], & m \in [10^{-6.5} ,\;  10^{-4.5}], \nn \\
U_0 \in [10^{-9.5} ,\;  10^{-6}],  &  \mu\in [0.01 ,\;  0.5],  &  \eta\in [0.5 ,\;  3.5].
\eqae
5 million parameter sets were investigated, of which 2,231,657 (44.6\%) satisfied the criteria for the background evolution\footnote{
The two most important criteria are that there should be a sufficient amount of
inflation to solve the horizon and flatness problems and
that the maximum value of $\gamma$ (considered over the entire run) should be greater than 1.
The latter case, of course, is far from unphysical as it corresponds to a two-field model with
slowly rolling coupled scalar fields. However, in this section we focus on deviations from the
standard scenario so exclude these models to focus on the more unusual phenomenology
associated with the DBI field.  
}. Most of the points rejected at this stage (60\%) yielded too few efolds, with the majority of the remainder exhibiting slow-roll in $\chi$ (38\%). 
With the range of parameters given above, the volume of the parameter space was
rather large, which meant that only a relatively small number of parameter sets (13063) were found that gave rise to 
a power spectrum amplitude within the observed range\footnote{
This can be understand by considering the analogous process in the single-field case, in 
which the power spectrum amplitude $P_{\rm amp}$ is proportional to the squared mass of the field. 
Increasing the volume of the parameter space (either by extending the range of mass values tested
or increasing the dimension of the parameter space) means that the number of cases that
{\it do not} give rise to the observational value of $P_{\rm amp}$ is vastly increased.
}.
This shall be referred to as the `better' set.
Similarly, the `best' set for which both the power spectrum and spectral index were within observational limits yielded 3834 points.

Except for some runs with the coupling $\beta$ very close to zero, no distinct region of the 
parameter space
was excluded by discounting the parameter sets that did not
satisfy the criteria for the background evolution  (i.e. gave rise to an extremely short
 inflationary stage or standard slow-roll models).
The relatively large range of parameters considered means that the calculated value of the power spectrum
amplitude
varies over several orders of magnitude with a strong dependence on the mass scale of the 
DBI field (as is to be expected since it is the perturbations of the DBI field that are the dominant contribution to
the curvature perturbation). Larger values of the $\varphi$-field potential, corresponding to small $\eta$ and large $U_0$, increase the Hubble parameter and also give rise to larger values of $P_{\rm amp}$.

\begin{figure}[t]
	\begin{center}
		\begin{tabular}{cc}
			   \includegraphics[width=0.5\textwidth]{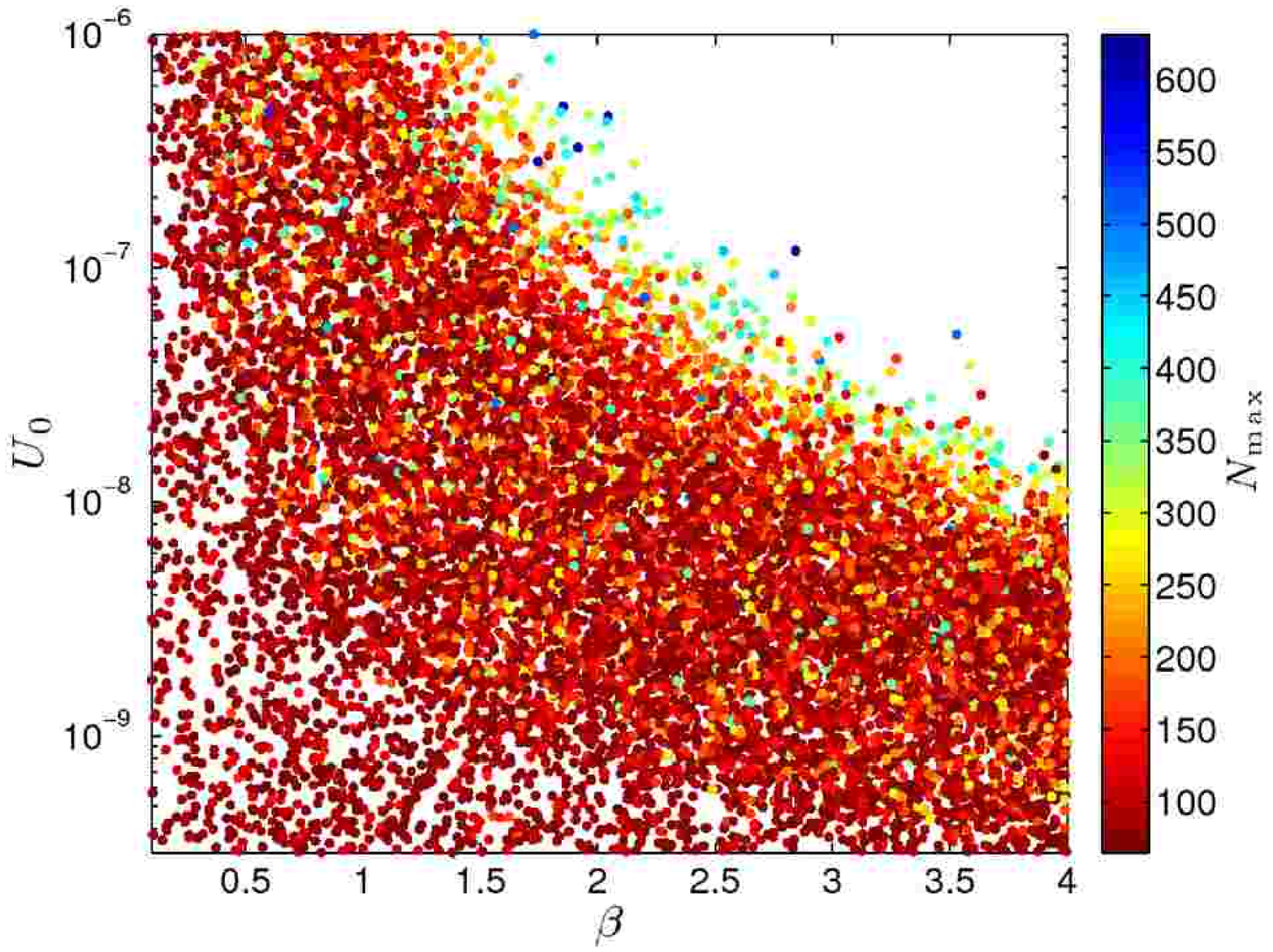} 
			& \includegraphics[width=0.5\textwidth]{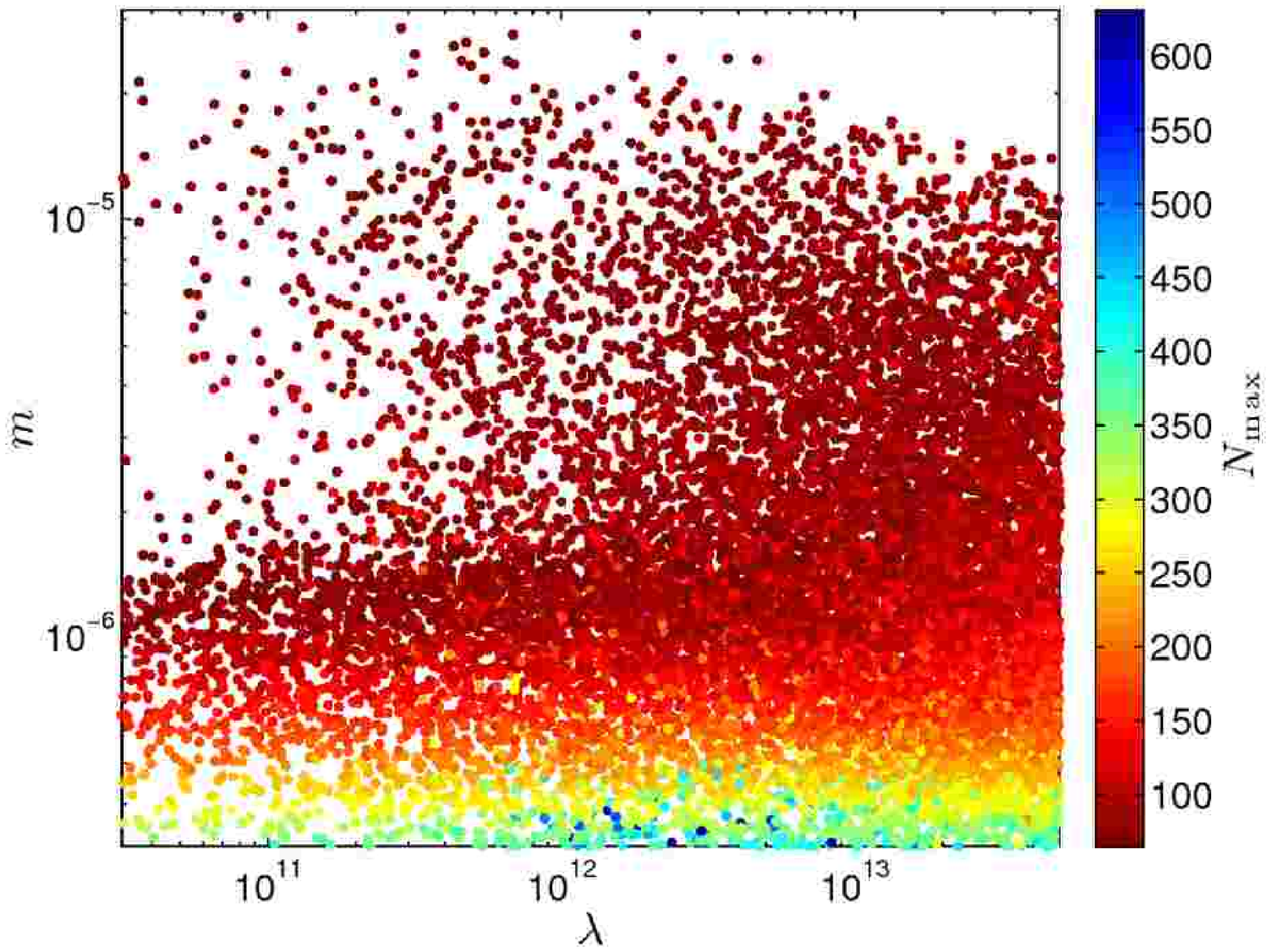} \\
			     \includegraphics[width=0.5\textwidth]{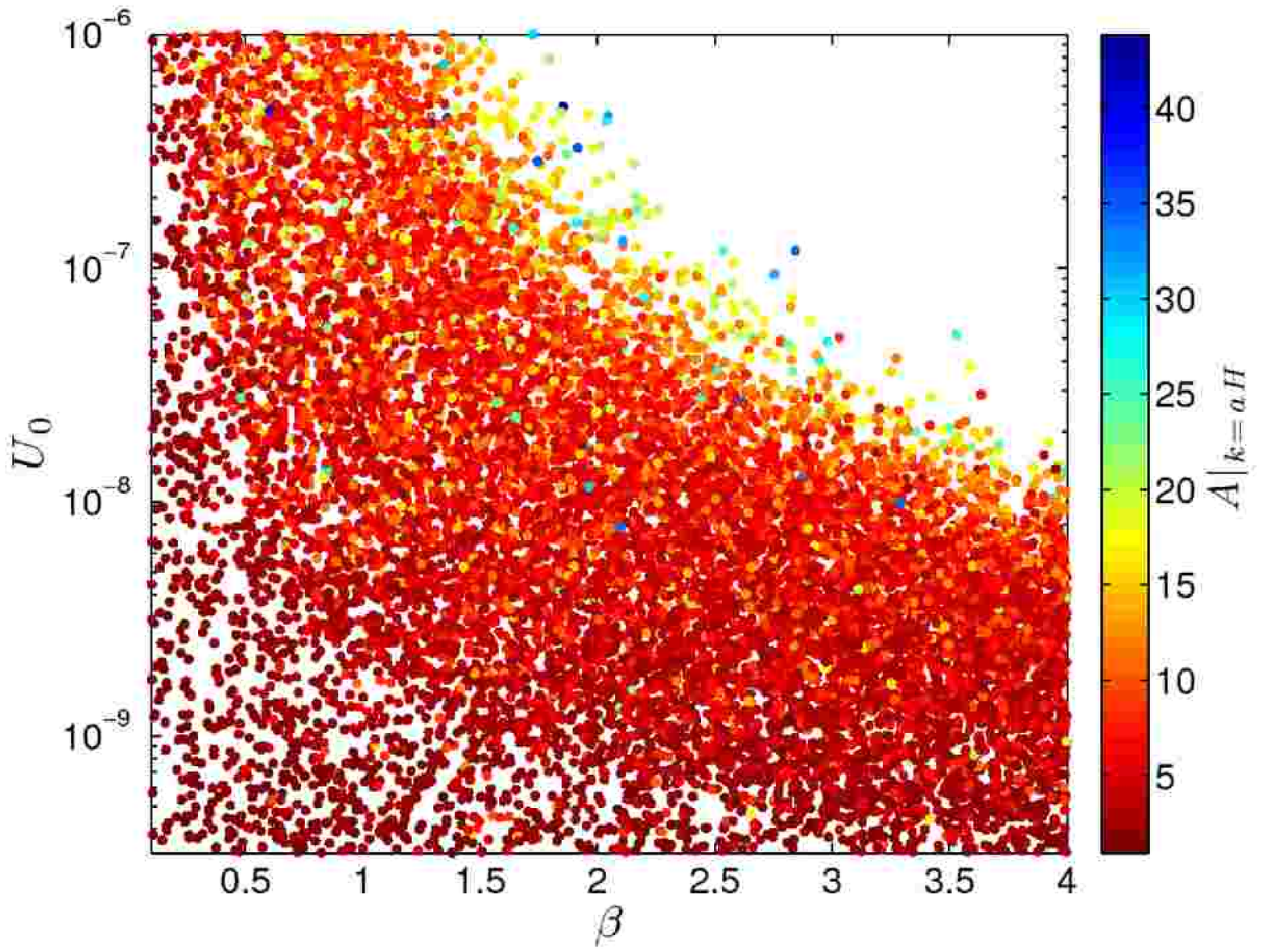} 
			& \includegraphics[width=0.5\textwidth]{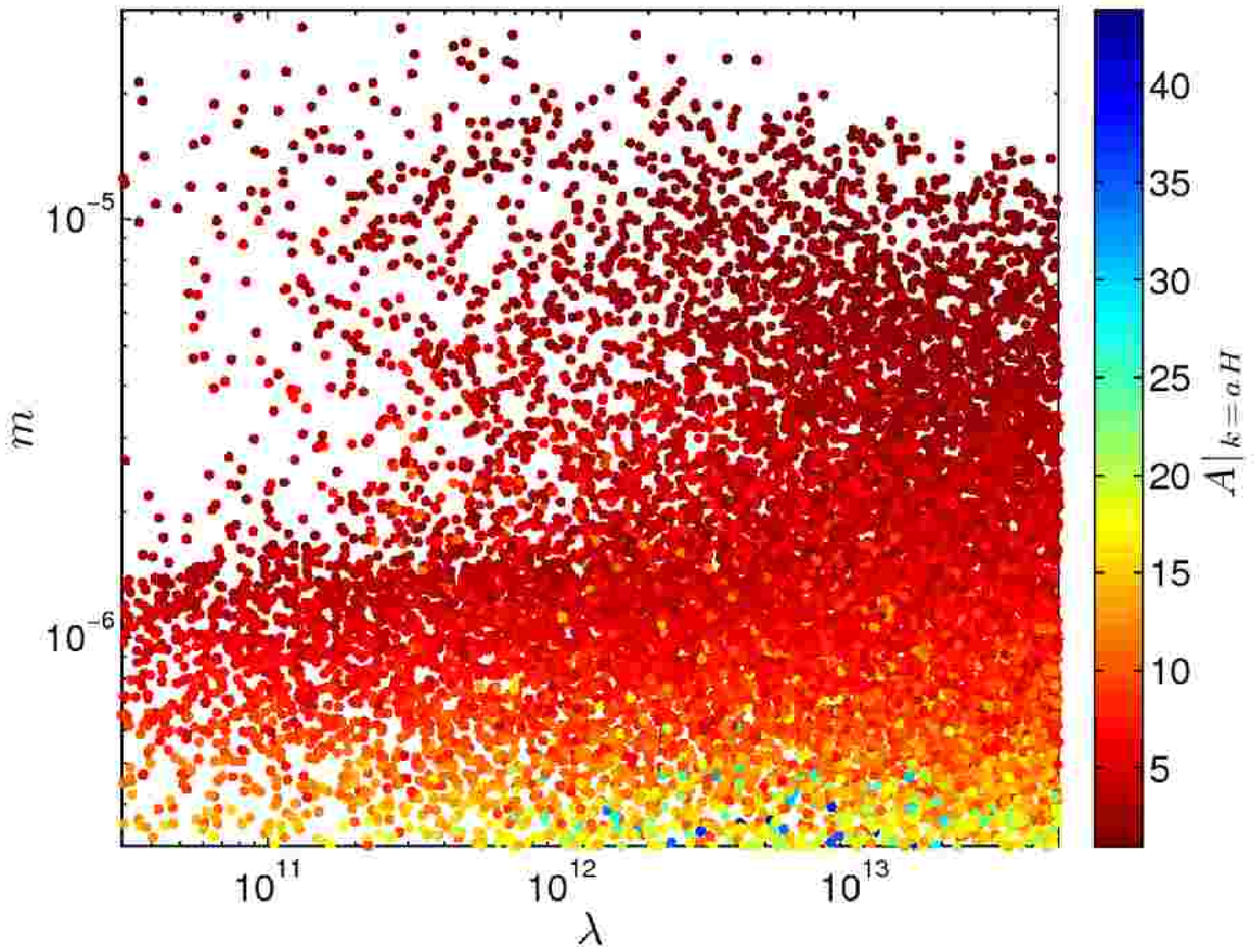} 
		\end{tabular}
	\end{center}
	\caption[The $\beta-\log_{10}U_0$ and $\log_{10}\lambda-\log_{10}m$ planes
	for the `better set' for the exponential potential.]
	{
	The $\beta-\log_{10}U_0$ (left column) and $\log_{10}\lambda-\log_{10}m$ (right column) planes
	for the `better set' for the exponential potential.
	The upper plots are colour-coded according to the number of efolds in the run $N_{\rm max}$
	and the lower plots according to the value of $A$ at $k_{\rm pivot}=aH$.
	 }
	\label{Exp_Better_ANk_NMax}
\end{figure}

\begin{figure} [!ht]
	\begin{center}
	 	\begin{tabular}{cc}
			   \includegraphics[width=0.45\textwidth]{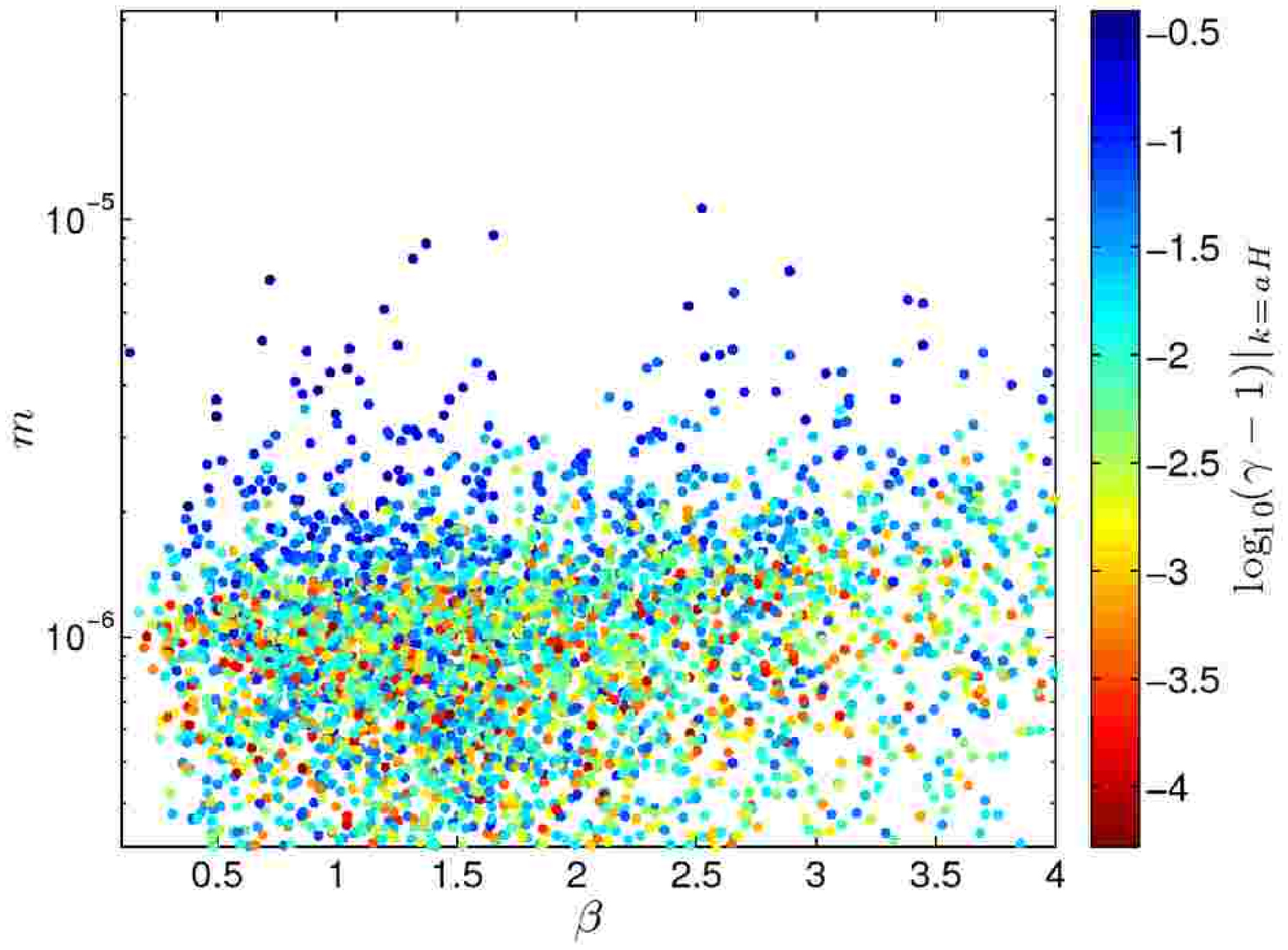} 
			& \includegraphics[width=0.45\textwidth]{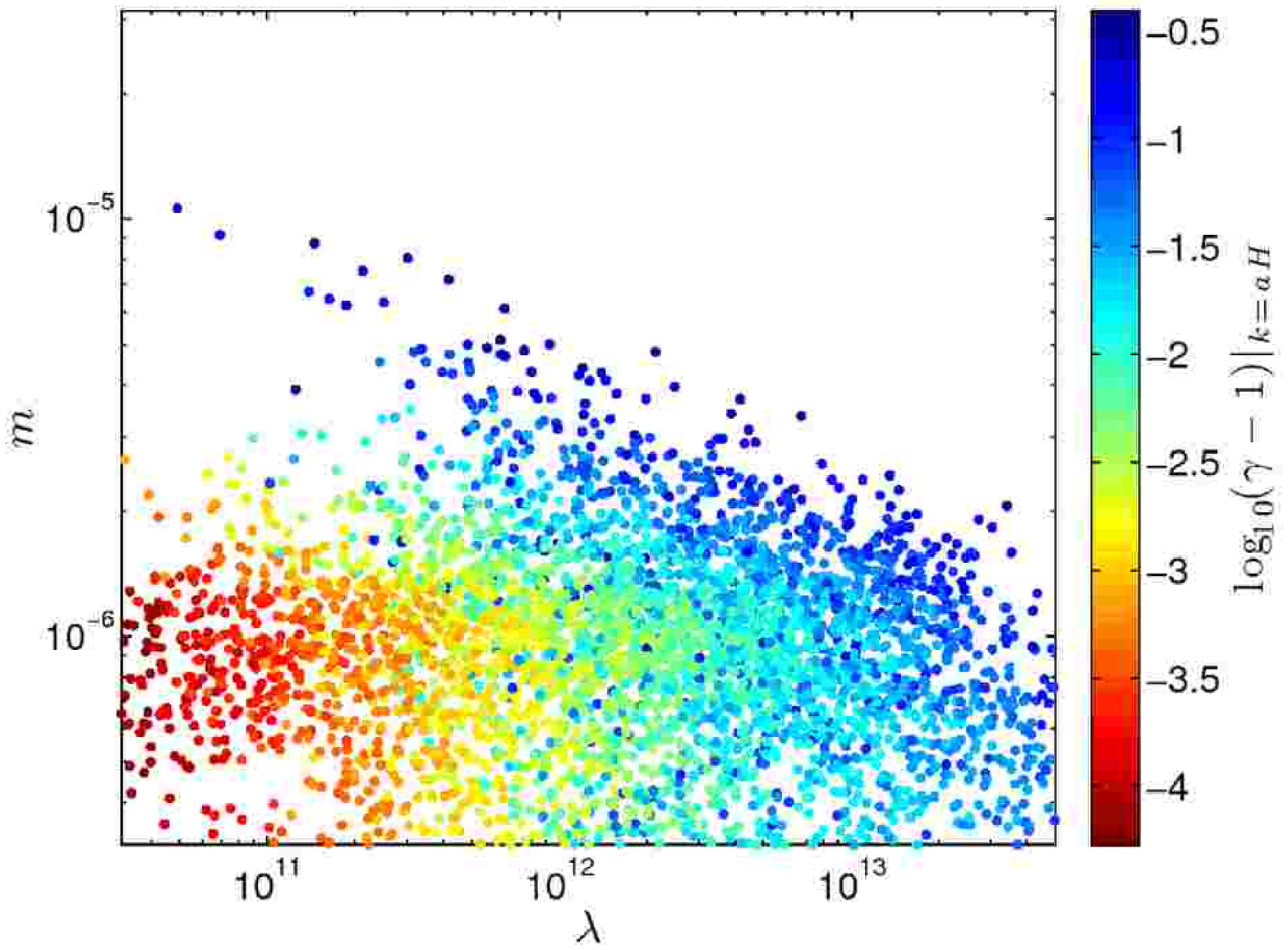} \\
			     \includegraphics[width=0.45\textwidth]{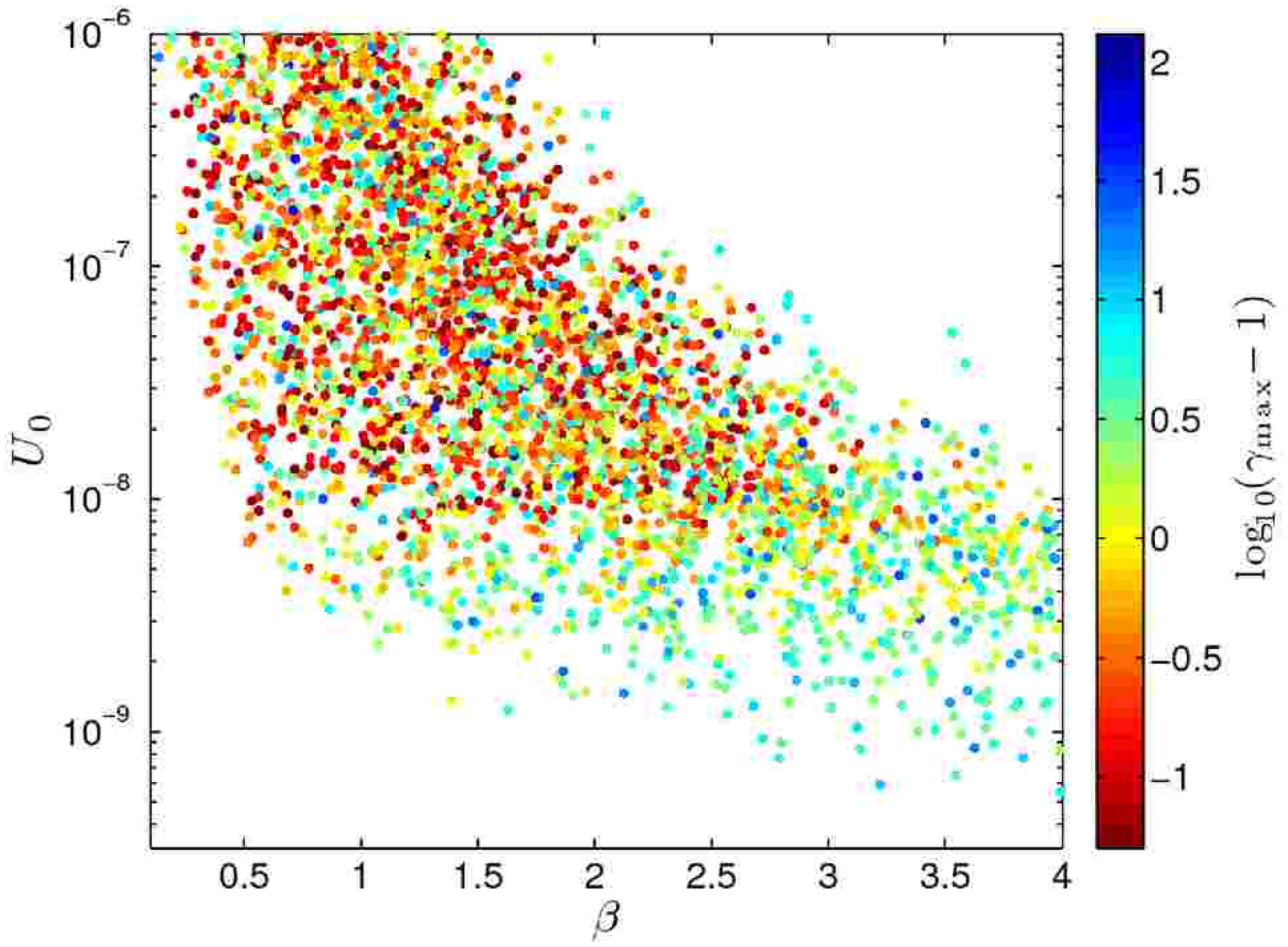} 
			& \includegraphics[width=0.45\textwidth]{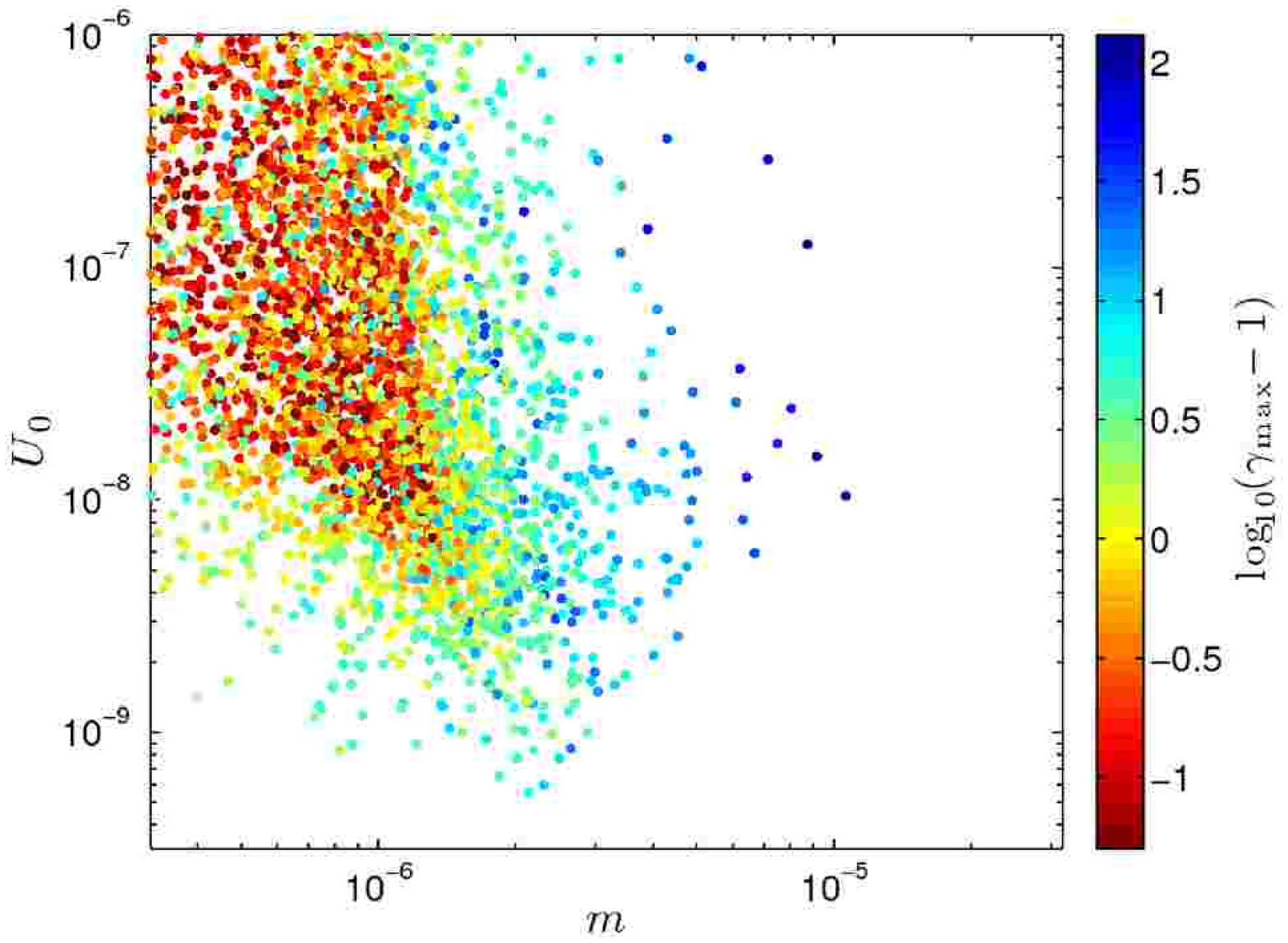} \\
			     \includegraphics[width=0.45\textwidth]{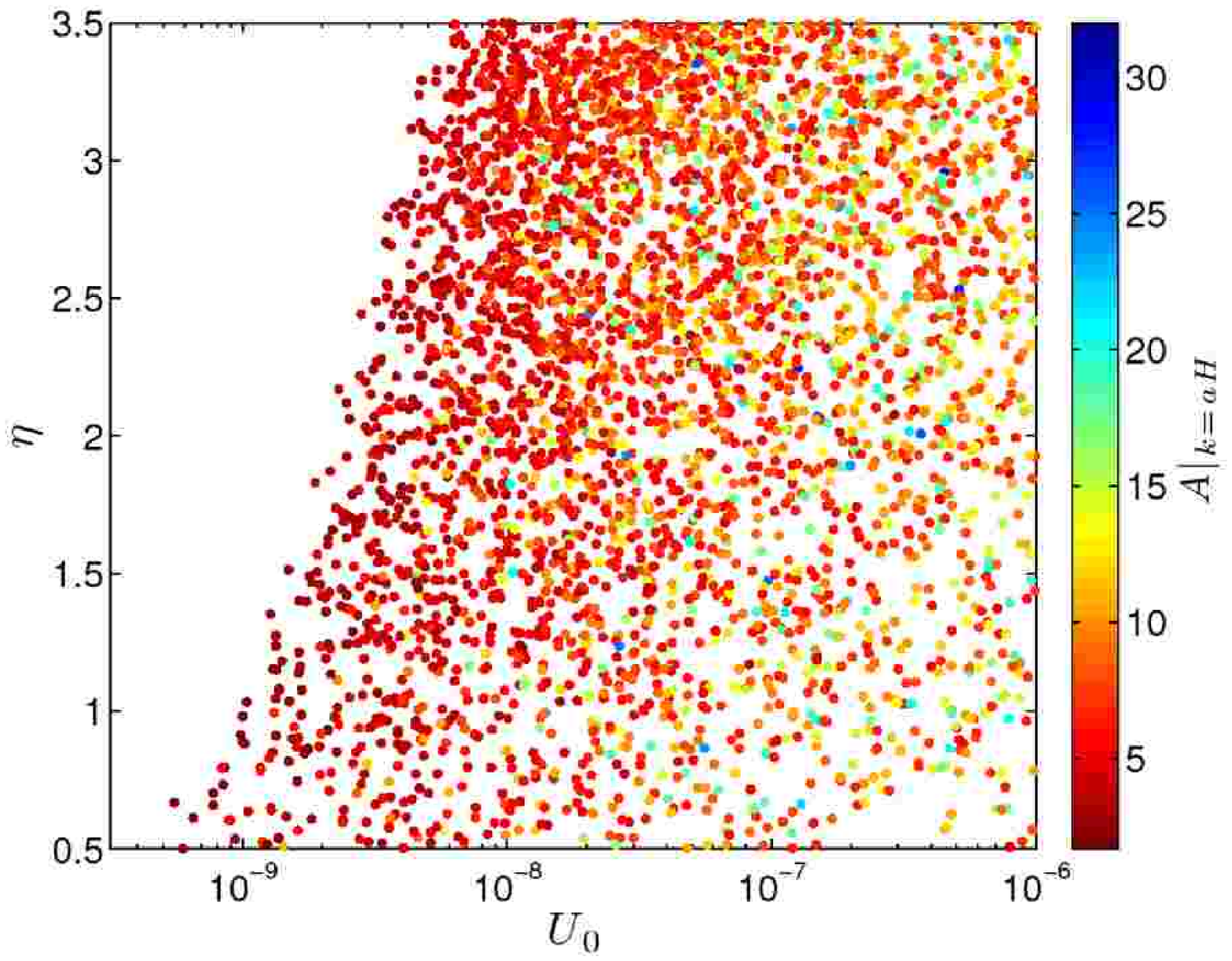} 
			& \includegraphics[width=0.45\textwidth]{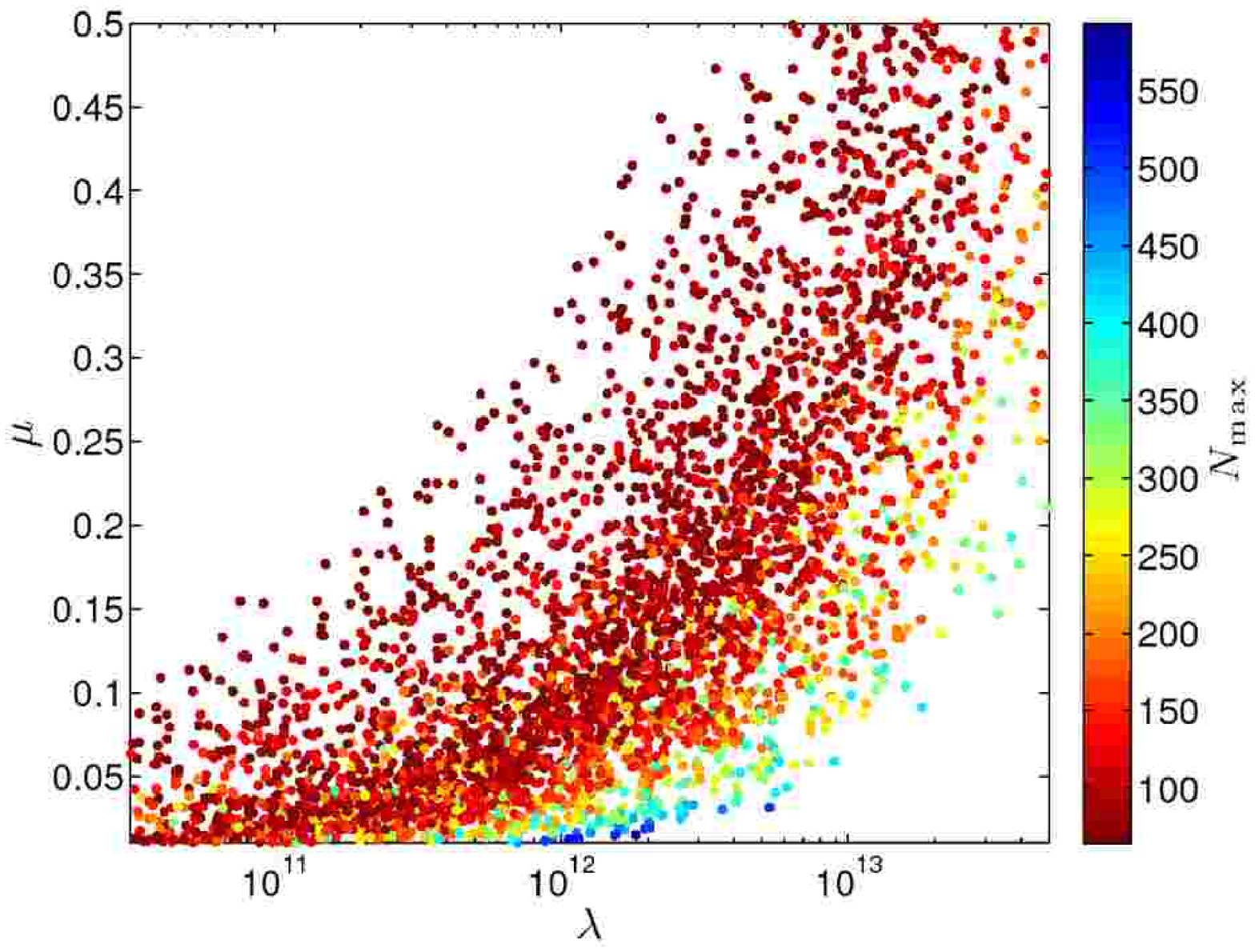}
		\end{tabular}
	\end{center}
	\caption[Views of the parameter space for the `best set' (for which the calculated values of
	$n_s$ and $P_{\rm amp}$ are within observational limits) for the exponential potential.]
	{
	Views of the parameter space for the `best set' (for which the calculated values of
	$n_s$ and $P_{\rm amp}$ are within observational limits) for the exponential potential.
	Top row: the $\beta-\log_{10}m$ (left) and $\log_{10}\lambda-\log_{10}m$ (right) planes,
	colour-coded according to the value of $\log_{10}(\gamma-1)$ at $k_{\rm pivot}=aH$.
	Middle row: the $\beta-\log_{10}U_0$ (left) and $\log_{10}m-\log_{10}U_0$ (right) planes,
	colour-coded according to the value of $\log_{10}(\gamma_{\rm max}-1)$.
	The bottom row shows
	 the $\log_{10}U_0-\eta$ plane, 
	colour-coded according to the value of $A$ at $k_{\rm pivot}=aH$,
	and the $\log_{10}\lambda-\mu$ plane, colour-coded according to the value of $N_{\rm max}$.
	}
	\label{Exp_Best}
\end{figure}

Focusing on the `better set' of runs with $P_{\rm amp}$ within observational limits, one can see
in fig. \ref{Exp_Better_ns} that the parameter space is significantly constrained compared to the `good set',
although the range of values of the spectral index $n_s$ is quite large. There appear to be two
overlapping regions in the plots in  fig. \ref{Exp_Better_ns}. The light blue points with $n_s\approx 0.97$ make up the bulk of the `best set' (shown in more detail in fig. \ref{Exp_Best}) and 
have $\gamma\simeq 1$ as the pivot scale crosses the horizon, so the spectral index is 
similar to that in slow-roll models
as DBI corrections are minimal.
To illustrate more clearly the qualitative difference in the properties of the background runs that
give rise to the distinct regions in fig.  \ref{Exp_Better_ns}, correlations between the boost factor and $n_s$
are shown in fig. \ref{Gam_ns}.
As $\gamma$ is a dynamical quantity that increases to a maximum as inflation proceeds, two views 
are shown: in the left panel, the value taken when the pivot mode leaves the horizon $(\gamma-1)|_{k_{\rm pivot = aH}}$ 
(corresponding to $N=55$) and the maximum value obtained 
throughout the background run $(\gamma_{\rm max}-1)$ (right panel).
As the value of $n_s$ shown on the vertical axis is a constant property for a given
run, the points can be thought of moving horizontally between the left panel and the right
as inflation procedes. 
It can be seen that the runs which do exhibit significant DBI characteristics (as indicated by $\gamma_{\rm max}\gtrsim 100$ in the right-panel plot) are precisely those with larger values of the boost factor at horizon crossing that are excluded by virtue of a blue-tilted spectral index.
This is, however, not to say that there are no viable scenarios that exhibit DBI characteristics as
some of the darker blue points 
--- which, as can be seen by comparing figs.  \ref{Exp_Better_ns} and \ref{Gam_ns},
are associated with smaller values of $\lambda$ and $\mu$, and a DBI mass scale $m\sim \mathcal{O}(10^{-6})$ ---
correspond to runs in which the boost factor increases from a very tiny value to  
$\gamma_{\rm max}\sim\mathcal{O}(10)$.
In both cases, deviations from the quasi-slow-roll solutions with $\gamma-1\ll 1$ throughout the background
evolution give rise to a corresponding shift in the resulting spectral index of the perturbations, leading to either a blue-tilted (for the points corresponding to solutions bearing close resemblance to standard DBI models)
or a more red-tilted spectrum (for the points in which DBI characteristics are suppressed by the
presence of the second field).

The parameters $\lambda$ and $m$ strongly affect the other background quantities: the duration of inflation $N_{\rm max}$ and the coupling $A$ (measured at $k_{\rm pivot}=aH$). The models with $A\gtrsim10$ (and
$P_{\rm amp}$ in the observed range) have small $m$ to compensate (cf. fig. \ref{Exp_Better_ANk_NMax}). In a similar way to the
case shown in fig. 9 of \cite{vandeBruck:2010yw} these models can exhibit slow-roll inflation in $\chi$ so that $N_{\rm max}$ is very large. As might be expected, models with large $A$ have $\beta>1$; the left column of 
fig. \ref{Exp_Better_ANk_NMax} shows a noticeable correlation between $\beta$ and $\log_{10}U_0$ for these models.

Looking at the `best' set for this model, with both $n_s$ and $P_{\rm amp}$ within observational limits (fig. \ref{Exp_Best}) it can be seen that most of the allowed models have very similar properties, with the boost factor $\gamma$ at horizon crossing deviating only slightly from $1$ and rising to values less than $\mathcal{O}(10)$. 
The allowed range of the DBI mass $m$ is relatively small (one order of magnitude) compared to $U_0$ and a large range of $\beta$ values are compatible with observations. The plots showing 
the $\beta-\log_{10}U_0$ and $\log_{10}U_0-\eta$ display clear structure, favouring larger values for the 
potential of the $\varphi$-field (when $\beta$ is not large) than in typical slow-roll scenarios, further emphasising the non-dynamical role of the field.
The bottom-right plot, showing the $\log_{10}\lambda-\mu$ plane is also tightly constrained, with values
that lead to large $f(\chi)$ excluded. As in the standard DBI scenario, a larger value  of the warp factor
is related to a large number of efolds of inflation. 

\section{Conclusions} 
\label{Sec:DBI:Conclusions}

In this article we have investigated in detail the consequences of 
the scalar-tensor model of DBI inflation introduced in \cite{vandeBruck:2010yw}, paying particular
attention to the effect of the different parameters on the value of the spectral index. 
When the coupling between the two fields is non-zero,
the $\varphi$-field (with a canonical kinetic term in its Lagrangian)
is forced into the minimum of its effective potential. Although
it does not play a
dynamical role, the field indirectly affects the DBI field $\chi$ and the perturbations  via the conformal coupling factor $A=\exp(\beta\varphi)$. 
With the $\varphi$-field in its minimum and working in the `slow-roll limit' in which the equation of motion for the DBI field is dominated by the
potential $V(\chi)$ (with effective mass $mA$) we have shown that,
with appropriate slow-roll parameters, the perturbation equations can be
put in a form similar to that describing perturbations in single-field k-inflation,
with an additional small correction due to the two-field nature of the
full system. In the limit that this can be neglected,  
the resulting estimate for the spectral index is
\[
n_s = 1-2s+2\delta-6\epsilon\left( 1+\tfrac{2\beta}{\eta} \right) ,
\]
which is superficially similar to the standard expressions for the single field DBI scenario. However, the
the $\epsilon$ and $\delta$ parameters are proportional to $A^{-2}\left( 1+4\beta/\eta \right)^{-2}$
and $A^{-2}\left( 1+4\beta/\eta \right)^{-1}$ respectively, so tend to be smaller than their counterparts
in the uncoupled case (with all other parameters unchanged). The values of the slow-roll parameters,
are dependent on the background evolution of the fields (which in turn are determined by quantities such as the mass of the DBI field) and the actual value of the $n_s$ can exhibit either a blue- or red-tilt.

Understanding the effect of the free parameters of the model on the observable quantities is
essential:
the numerical work described in Sec. \ref{Sec:DbiParameterSpace}
represents a first step toward this end.
It is shown that the parameter space
can be severely constrained by the requirement that the mechanism must be capable of
driving a sufficient amount of inflation and by considering the values of the power spectrum amplitude and
the spectral index. The latter measurement in particular rules out a large class of models ---
those with small values of $c_s$ at horizon crossing, which give
rise to a blue-tilted spectrum ---
so that only scenarios in which DBI characteristics are suppressed at the beginning of the 
observable period of inflation are allowed. In general, the excluded models with $(\gamma-1)\gtrsim \mathcal{O}(1)$ at horizon crossing exhibit significant DBI effects (i.e. large values of the boost factor)throughout the inflationary period.
The remaining realisations of the model have $(\gamma-1) \ll 1$ at horizon crossing: some of which differ only slightly from standard slow-roll solutions. This is in general agreement with the results of other numerical
studies investigating the standard DBI model \cite{Bean:2007hc}
where it was found that only models exhibiting small DBI effects were viable.
However, a number of viable cases --- associated with a more red-tilted spectral index ---  were found for which the boost factor is initially suppressed by the effect of the coupling between the fields, but increases later to moderate values $\sim\mathcal{O}(10)$. 
Further work is needed to fully understand the viability of the model; a natural extension
is to consider other observables: in particular, the tensor-to-scalar ratio $r$ and the level of non-Gaussianity (proportional to $\gamma^2$ in the single-field case)
which so tightly constrains the standard DBI scenario. It would also be of interest to calculate the running of the spectral index, as including this gives different bounds on the spectral index \cite{Komatsu:2010fb}.

In conclusion, the scalar-tensor DBI inflation scenario provides a simple mechanism to reduce the
large values of the boost factor associated with single field models with DBI action, whilst still being able to drive 60 efolds of inflation. We have indicated the regions of the parameter space of the model capable of giving rise to a power spectrum with amplitude and spectral index within the observed bounds, and moreover shown analytically that the value of the latter quantity arises when
the coupling forces one field into its minimum, leaving the DBI field to slow-roll.

\begin{acknowledgments}
 D.F.M. thanks the Research Council
of Norway FRINAT grant 197251/V30. D.F.M. is also partially
supported by project PTDC/FIS/111725/2009 and CERN/FP/116398/2010.
J.M.W. was supported by Yggdrasil Grant 195757/V11 from the research council of Norway, and would like to thank the members of the Institute of Theoretical Astrophysics at the University of Norway for their kind hospitality while this work was initiated. We would also like to thank the anonymous referee for helpful and constructive comments. 

\end{acknowledgments}

\appendix

\section{Approximating $C_{\chi\chi}$ to first order in slow-roll parameters}
\label{DerivationAppendix}

In this appendix, we derive an first-order expression for the mass term $C_{\chi\chi}$ [given in full by
(\ref{CXX})] that appears in the perturbed equation of motion (\ref{First_uk_eqn}).
Before rewriting  each contribution to this equation in terms of the slow roll parameters $\epsilon$, $\delta$ and $s$, it is necessary to obtain an expression involving $f_\chi$. To do this, one can
differentiate $(1-c_s^2) = A^{-2}f\dot\chi^2$ to get a relation between $\ddot\chi$ and $f_\chi$. This yields
\eqa
\frac{\ddot\chi}{\dot\chi} &=& \beta\dot\varphi -\tfrac{1}{2}\frac{f_\chi}{f}\dot\chi
+\left(1-c_s^{-2}\right)^{-1}\frac{\dot{c_s}}{c_s}, \\
\Rightarrow \frac{\ddot\chi}{H\dot\chi} &\simeq& 
\frac{2\beta}{\eta}\epsilon -\tfrac{1}{2}\frac{f_\chi}{f}\frac{\dot\chi}{H}
+\left(1-c_s^{-2}\right)^{-1}s. \label{ddotchiApprox}
\eqae
Then differentiate $-2\dot H \simeq A^2\gamma\dot\chi^2$ to obtain,
\eqa
2\dot\epsilon H^2 + 4\epsilon H\dot H &\simeq& \left( 2\beta\dot\varphi -\dot{c_s}/c_s \right)(2\epsilon H^2)
+2A^2\gamma\dot\chi\ddot\chi, \nn\\
\frac{\dot\epsilon}{2H\epsilon}-\epsilon  &\simeq&
\frac{2\beta}{\eta}\epsilon-\tfrac{1}{2}s + \frac{\ddot\chi}{H\dot\chi}.
\eqae
Substituting (\ref{dotEps}) and (\ref{ddotchiApprox}) gives
\eq \label{fchiApprox}
 -\tfrac{1}{2}\frac{f_\chi}{f}\frac{\dot\chi}{H} \simeq
 \left(1-c_s^{2}\right)^{-1}s - \delta +\left( 1+\tfrac{2\beta}{\eta} \right)\epsilon.
\eqe
We can now write each contribution to $C_{\chi\chi}$ in (\ref{CXX}) in terms of the slow roll parameters $\epsilon$, $\delta$ and $s$, working to $\cal{O}(\epsilon)$ and dropping second order terms. Starting at the beginning
\begin{itemize}

\item The first term is
\eqa
\frac{A^4\dot\chi}{ H}\frac{f_\chi}{f^2}(1-c_s)^2 &=& -2A^2\dot\chi^2\left(\frac{1-c_s}{1+c_s}\right)
\left[  -\tfrac{1}{2}\frac{f_\chi}{f}\frac{\dot\chi}{H} \right], \nn\\
 &\simeq& -4c_s\epsilon H^2 \left(\frac{1-c_s}{1+c_s}\right)
\left[  -\tfrac{1}{2}\frac{f_\chi}{f}\frac{\dot\chi}{H} \right], \nn\\
&\simeq&
-4c_s\epsilon H^2\left(\frac{1-c_s}{1+c_s}\right)\left[  \left(1-c_s^{2}\right)^{-1}s - \delta +\left( 1+\tfrac{2\beta}{\eta} \right)\epsilon \right], \nn\\
&\sim& {\cal O}(\epsilon^2 H^2). \nn
\eqae
using $A^2\gamma\dot\chi^2 \simeq 2\epsilon H^2$.

\item The second term is
\eqa
-\left[ \frac{f_\chi}{f} -\frac{A^2\dot\chi}{ H c_s}\right]\frac{\dot{c_s}}{c_s}\dot\chi
&=&
\left[ -\frac{f_\chi}{f}\frac{\dot\chi}{H} + \frac{A^2\gamma\dot\chi^2}{H^2}\right]
\left( \frac{\dot{c_s}}{H c_s} \right) H^2, \nn\\
&\simeq&
\left[  2\left(1-c_s^{2}\right)^{-1}s - 2\delta +4\left( 1+\tfrac{\beta}{\eta} \right)\epsilon \right]sH^2, \nn\\
&\sim& {\cal O}(\epsilon^2 H^2).\nn
\eqae

\item The third term is
\eqa
-\tfrac{1}{2}c_s f_\chi A^{-4} \dot\chi^2 V_{T,\chi}   
&=&
 -\tfrac{1}{2}c_s f_\chi \dot\chi^2 V_\chi \nn\\
&=&
 \tfrac{1}{2} \frac{f_\chi}{Hf}(1-c_s^2) (-A^2 c_s V_\chi)H \nn\\
&\simeq&
3H^2(1-c_s^2) \left[ \tfrac{1}{2} \frac{f_\chi\dot\chi}{Hf} \right] \nn\\
&\simeq&
3H^2\left[ (1-c_s^2)\delta -s - (1-c_s^2)\left(1+\tfrac{2\beta}{\eta}\right)\epsilon \right]
\label{term3}
\eqae
where (\ref{ChiApprox}) was used in the third step and (\ref{fchiApprox}) in the last.

\item The fourth term is
\eqa
\tfrac{1}{2}A^2 (1-c_s)^2 \left[ c_s \left(\frac{f_\chi}{f^2}\right)_{,\chi} +(1+c_s)f^{-1}\left(\frac{f_\chi}{f}\right)_{,\chi} \right]
=
\frac{1-c_s}{2(1+c_s)}\left[ (1+2c_s)-c_s  \left(\frac{f_\chi}{f}\right)\right] \left(\frac{f_\chi}{f}\right)_{,\chi}\dot\chi^2\nn.
\eqae
The final part of this term can be expressed in terms of
time derivatives of the slow-roll parameters and other second-order terms
\eqa
\tfrac{1}{2}\left(\frac{f_\chi}{f}\right)_{,\chi}\dot\chi^2 &=& 
\left( \frac{f_\chi\dot\chi}{2f}  \right)\dot{\vphantom{\big| }}-\tfrac{1}{2}\frac{f_\chi}{f}\ddot\chi, \nn\\
&=&
H\left( \frac{f_\chi\dot\chi}{2Hf}  \right)\dot{\vphantom{\big| }} + H^2 \left( \frac{f_\chi\dot\chi}{2Hf}  \right)\left(  \epsilon - \frac{\ddot \chi}{H\dot \chi} \right) \nn\\
&\sim& \mathcal{O}(H^2\epsilon^2),
\eqae
using (\ref{fchiApprox}) and (\ref{ddotchiApprox}).  We have checked numerically to
 confirm that the contribution of this term to $C_{\chi\chi}$
 is negligible.

\item The fifth term is
\eq
\tfrac{3}{2}A^2\dot\chi^2 c_s^{-1}(1+c_s^2) \simeq 3H^2(1+c_s^2)\epsilon. \label{term5}
\eqe

\item The sixth term is
\eq
-A^4 c_s^{-2}\frac{\dot\chi^4}{2 H^2} \simeq 2H^2\epsilon^2 \sim {\cal O}(\epsilon^2 H^2).\nn
\eqe

\item The seventh term is
\eq
-A^2 c_s^{-1}(1+c_s^2)\frac{\dot\chi^2\dot{\varphi}^2}{4 H^2}
\simeq -\frac{2}{\eta^2}(1+c_s^2)H^2\epsilon^3 \sim {\cal O}(\epsilon^3 H^2).\nn
\eqe
using (\ref{SecondPhiApprox}).

\item The eighth term is
\eqa
\frac{\dot\chi V_{T,\chi}}{ H}(1+c_s^2) &\simeq&
-(1+c_s^2)\frac{A^6 V_\chi^2}{3\gamma H^2}, \nn\\
&\simeq&
-3H^2(1+c_s^2)\left[ \frac{1}{A^2\gamma} \left( 1+\tfrac{4\beta}{\eta} \right)^{-2}\frac{V_\chi^2}{V^2} \right], \nn\\
&\simeq&
-6H^2(1+c_s^2)\epsilon. \label{term8}
\eqae

\item The final term is
\eqa
c_s^{3}A^{-2} V_{T,\chi\chi} &=& c_s^{3}A^2 V_{\chi\chi}, \nn\\
&=& 3H^2 c_s^{2} \left[\frac{1}{A^2\gamma}  \left( 1+\tfrac{4\beta}{\eta} \right)^{-1}\frac{V_{\chi\chi}}{V} \right], \nn\\
&=& 3H^2c_s^2\delta. \label{term9}
\eqae

\end{itemize}

Combining eqns. (\ref{term3}),  (\ref{term5}),  (\ref{term8}) and  (\ref{term9}) then gives eqn. (\ref{CChiChi})
\[
C_{\chi\chi} \simeq 3H^2\left( \delta -s -2\epsilon\left[1+\tfrac{\beta}{\eta} \left( 1-c_s^2 \right) \right]\right).
\]


\section{Comparing the exponential and quadratic potentials}
\label{QuadraticPotentialAppendix}

\begin{figure}
	\begin{center}
		\begin{tabular}{c}
		\includegraphics[width=0.9\textwidth]{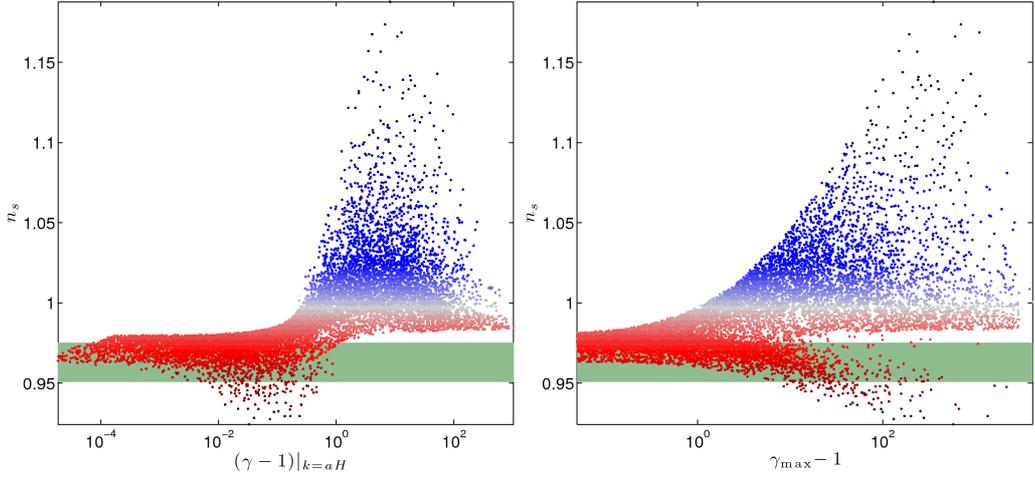}
		\end{tabular}
	\end{center}
		\caption{
		The distribution of $(\gamma-1)$ values for the offset quadratic potential. Quantities plotted
		are the same as in fig. \protect\ref{Gam_ns}.
	 }
	 \label{Gam_ns_Quad}
\end{figure}

\begin{figure}
	\begin{center}
	 	\begin{tabular}{cc}
			   \includegraphics[width=0.45\textwidth]{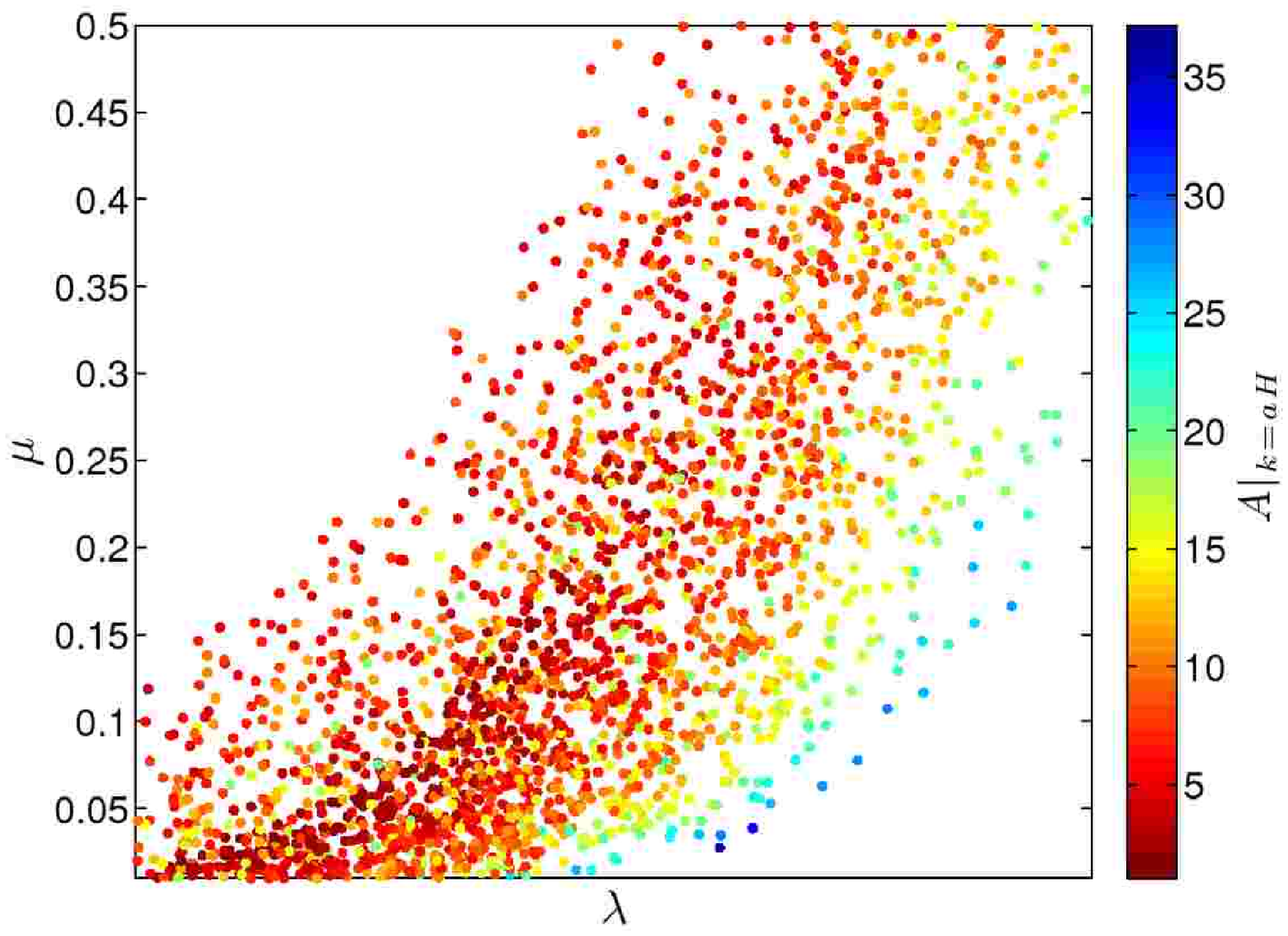} 
			& \includegraphics[width=0.45\textwidth]{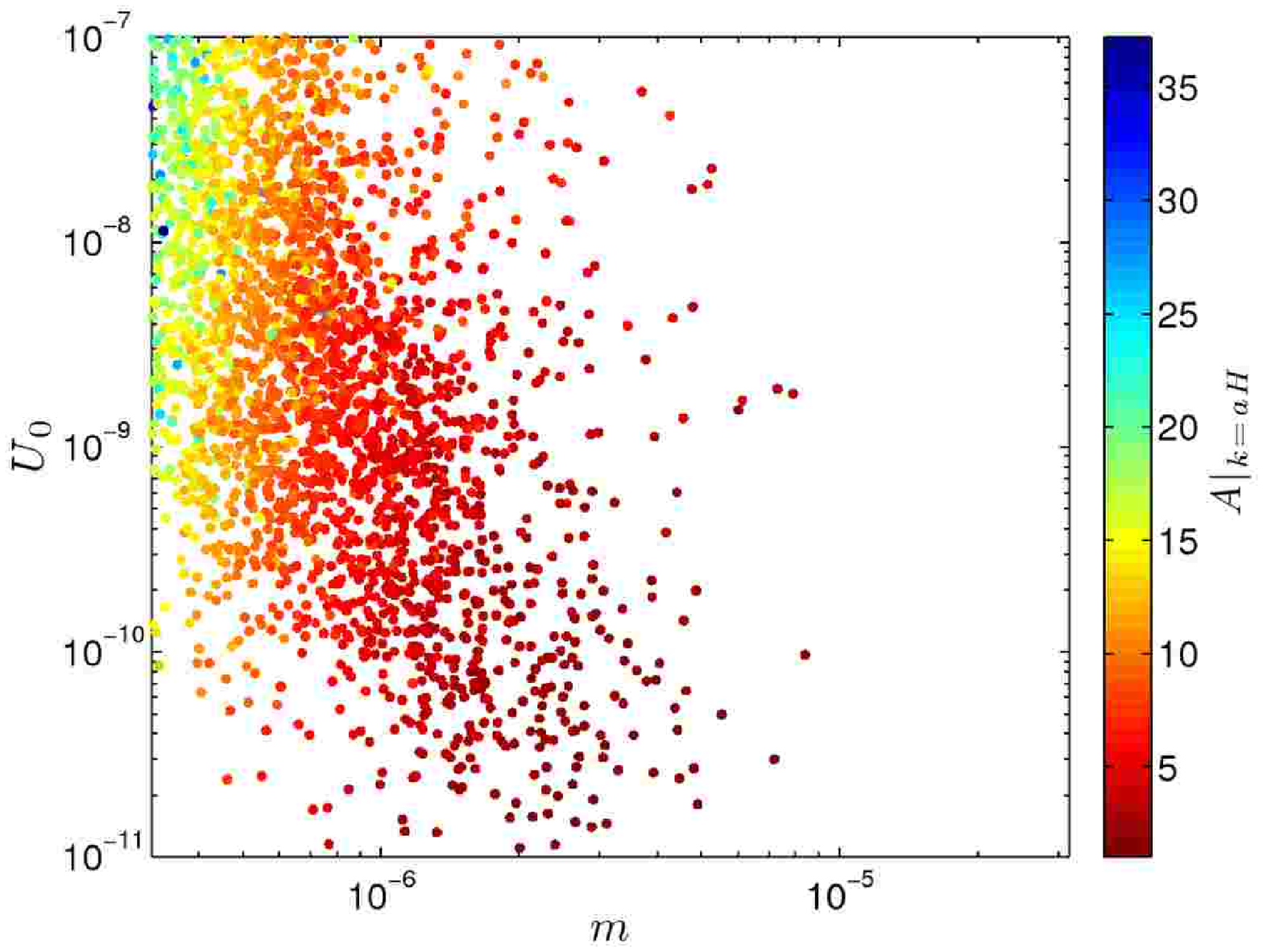}
		\end{tabular}
	\end{center}
	\caption{
	Two views of the parameter space for the `best set' (for which the calculated values of
	$n_s$ and $P_{\rm amp}$ are within observational limits) for the quadratic potential,
	colour-coded according to the value of $A$ at $k_{\rm pivot}=aH$.
	Left panel: $\log_{10}\lambda-\mu$; right panel: $\log_{10}m-\log_{10}U_0$.
	}
	\label{Quad_Best_Ak}
\end{figure}

As mentioned in the body of the article, the numerical analysis
in Sec. \ref{Sec:DbiParameterSpace} was repeated with an offset 
quadratic potential for $\varphi$-field (as described in \cite{vandeBruck:2010yw})
given by
\eq \label{DBI:OffQuad}
U(\varphi) = U_0 (\varphi-\eta)^2.
\eqe
Like the exponential potential, this is dependent on 
two free parameters $U_0$ and $\eta$: in this case
the latter gives the value of the true minimum is $\varphi =  \eta$.
The parameter ranges were
\eqa
\beta \in [0.1 ,\; 4],  & \lambda \in [10^{10.5} ,\;  5\times10^{14}], & m \in [10^{-6.5} ,\;  10^{-4.5}], \nn \\
U_0 \in [10^{-11} ,\;  10^{-7}],  &  \mu\in [0.01 ,\;  0.5],  &  \eta\in [1 ,\;  6].
\eqae
Again, 5 million parameter sets were investigated, of which 2,748,236 (55.0\%) satisfied the criteria for the background evolution. As with the exponential potential, of the points rejected at this stage, most (56\%) yielded too few efolds, with the majority of the remainder exhibiting slow-roll in $\chi$ (41\%). 11,999 of the acceptable parameter sets had a power spectrum amplitude within the observed range, and the `best' set (for which both the power spectrum and spectral index were within observational limits) yielded 2548 points.

The parameter space for the quadratic potential is very similar to that of the exponential potential in many respects, with the exception of those involving the parameter $\eta$, which has a different interpretation in each case. For example, for the exponential potential smaller values of $\eta$ give rise to larger values of the power spectrum amplitude, due to the increase in size of the potential $U(\varphi)$. For the quadratic potential, for the same reason, it is larger values of $\eta$ that are correlated with $P_{\rm amp}$.  Focusing on the parameter space for the `better' set for this potential 
 consisting of parameters that give rise to $P_{\rm amp}$
within observational limits, it is again possible to identify two groups of points in the parameter space.
The interpretation is the same as in the case of the exponential potential: 
runs that deviate significantly from quasi-slow-roll behaviour (i.e. $(\gamma-1)\ll 1$) as the pivot mode leaves
the horizon give rise to a blue-tilted spectrum with $n_s>1$ (blue points)
and runs in which DBI characteristics are suppressed at $N=55$, but become significant later 
during inflation, exhibit a strong red-tilt (dark red points).
The similarity between
the two potentials can be seen by comparing figs. \ref{Gam_ns} and \ref{Gam_ns_Quad},
which show the correlation between the spectral index and the boost factor.

Looking at the `best' set, with both the power spectrum amplitude and spectral index satisfying observational constraints, 
we see more similarities between the results for the quadratic and exponential potentials. The distribution of points in the views of the parameter space showing the mass parameters and the warp factor parameters shown in fig. 
\ref{Quad_Best_Ak} are almost identical to those for the exponential potential (cf. middle-right and bottom-right plots in fig. \ref{Exp_Best}). Again, there is a strong correlation between
small values of the DBI mass and
large values of the coupling $A$ (and also large values of $N_{\rm max}$).

\begin{figure}
	\begin{center}
	 	\begin{tabular}{cc}
			\includegraphics[width=0.5\textwidth]{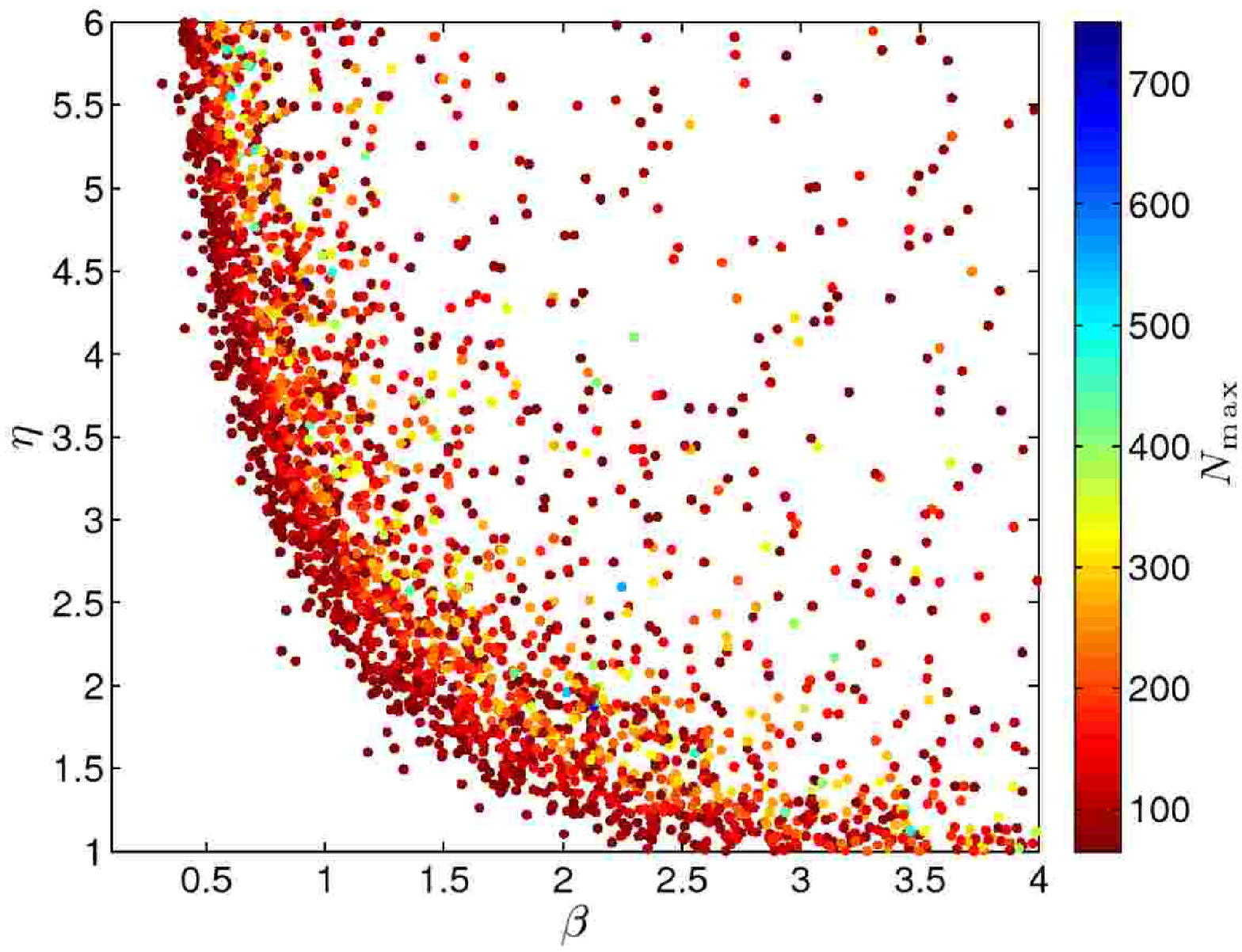} 
			& \includegraphics[width=0.5\textwidth]{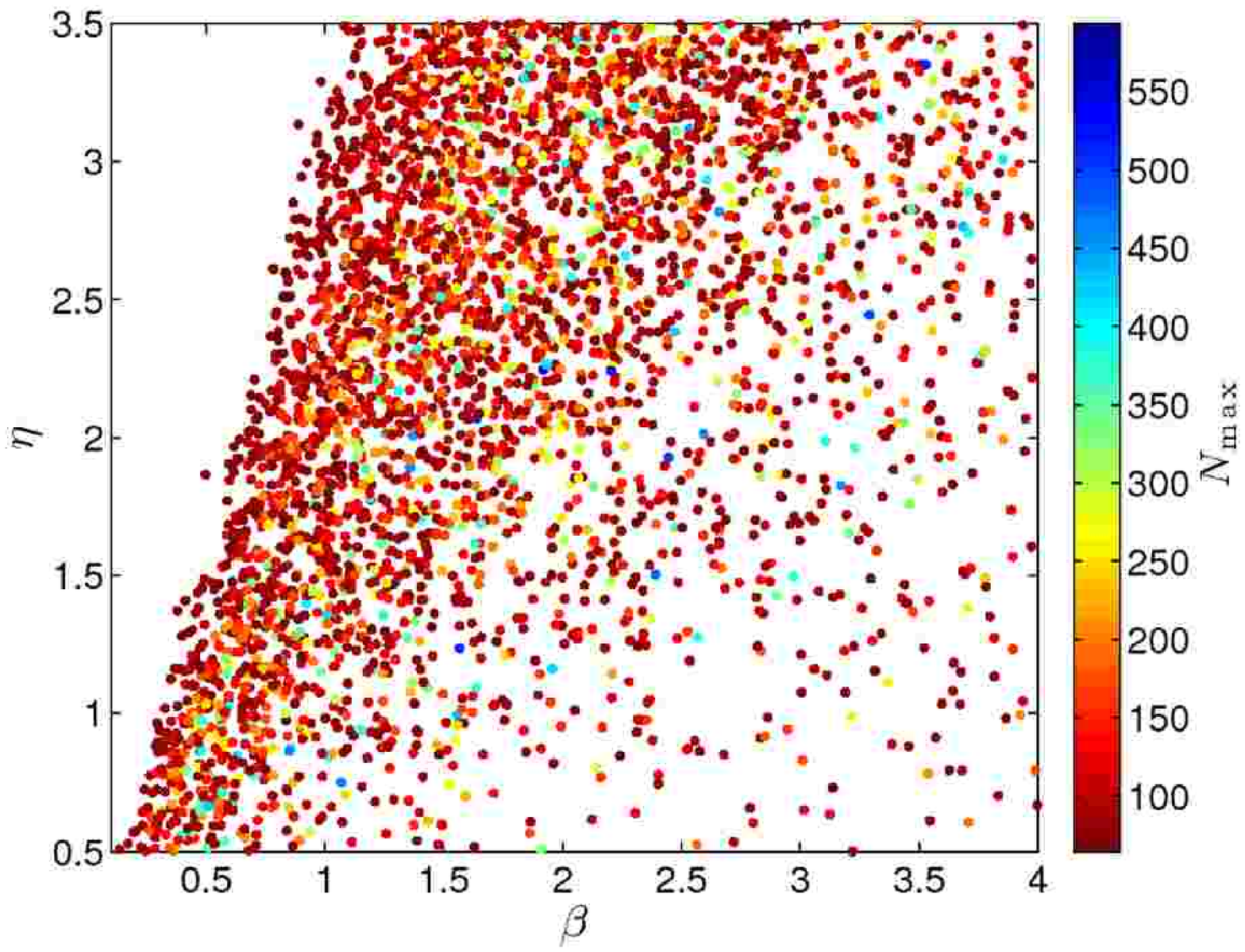} 
		\end{tabular}
	\end{center}
	\caption{
	The $\beta-\eta$ planes for the quadratic potential (left) and the exponential potential (right), 
	using the `best sets'. Both plots are colour-
	coded according to the value of $N_{\rm max}$ (cf. analytical constraints shown in fig. 
	\protect\ref{Best_ExpQuadCompare}).
	}
	\label{eta_beta_plots}
\end{figure}

\begin{figure}
	\begin{center}
		\begin{tabular}{cc}
			 \includegraphics[width=0.5\textwidth]{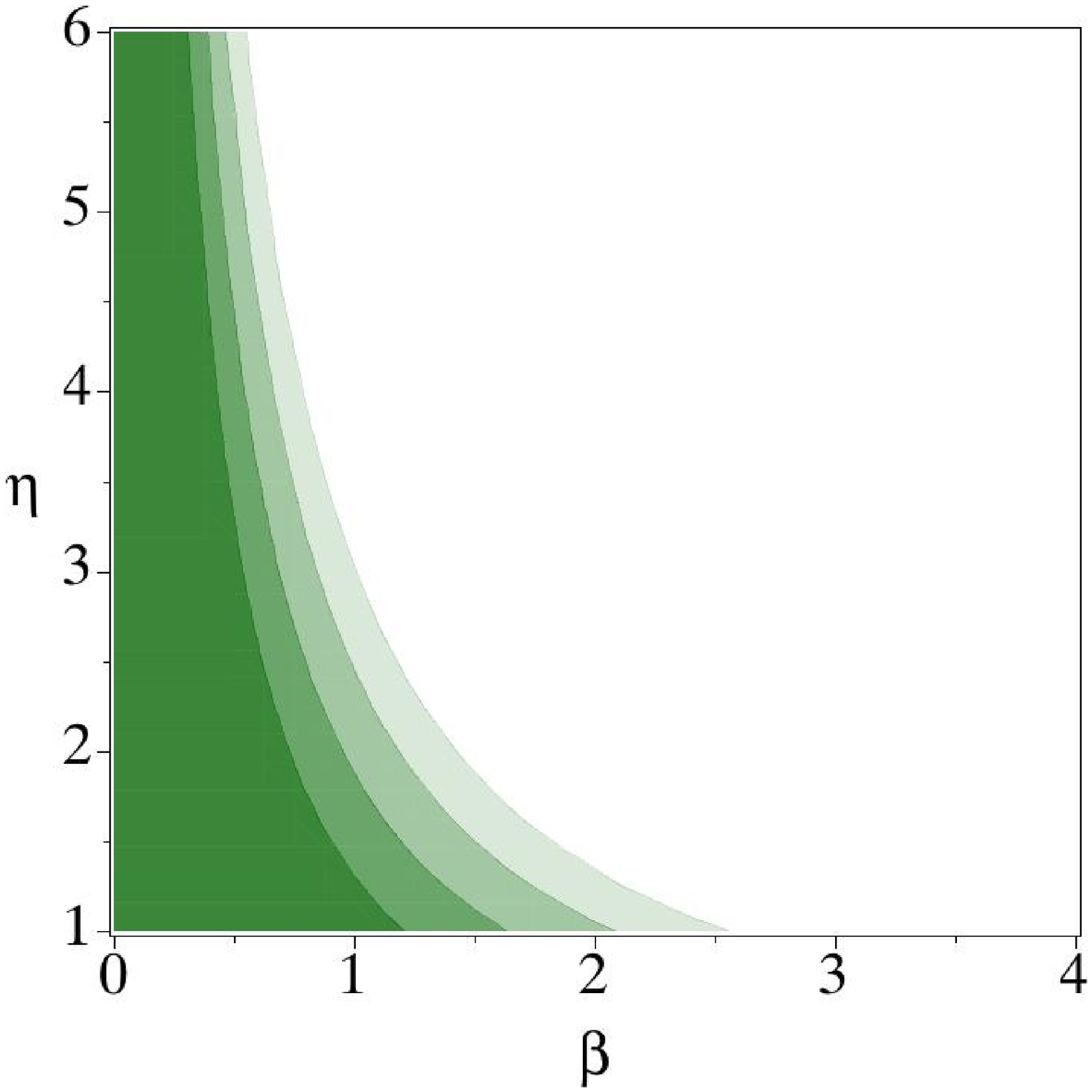} 
			& \includegraphics[width=0.5\textwidth]{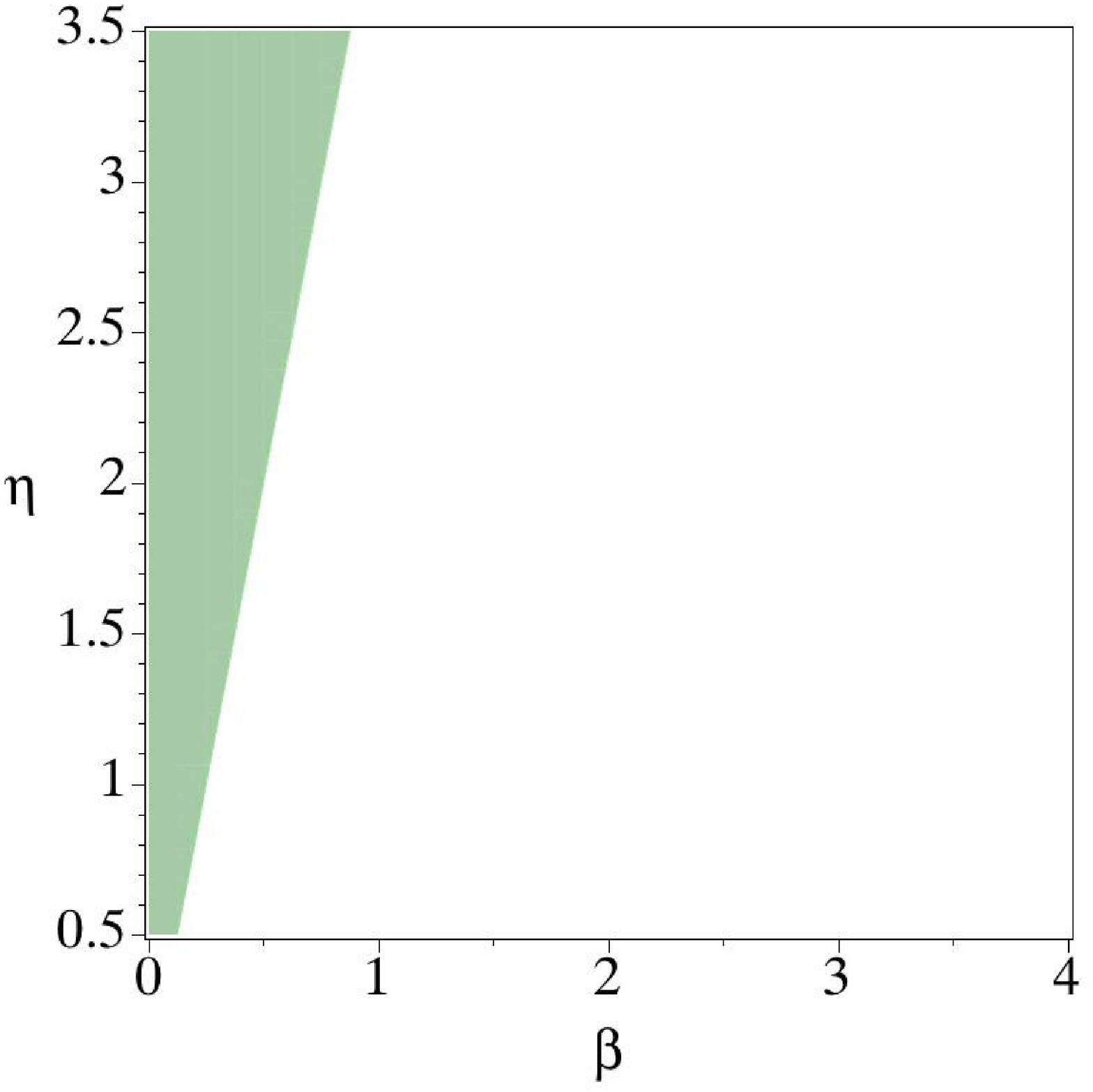} 
		\end{tabular}
	\end{center}
	\caption{
	For comparison with fig. \protect\ref{eta_beta_plots},
	the shaded regions show the regions of the $\eta-\beta$ planes in which the
	condition  $\rho_{\rm DBI}>\rho_{\varphi}$ is satisfied for the quadratic (left panel)
	and exponential
	potentials (right panel) using the same parameter ranges.
	The former condition is dependent on the ratio $V/U_0$ (see text)
	so we have used $V/U_0 = 10^{-5}$, $10^{-4}$, $10^{-3}$, $10^{-2}$, with
	the lightest shade corresponding to $10^{-5}$.
	}
	\label{Best_ExpQuadCompare}
\end{figure}

The general distribution of the points in the parameter space is largely the same for both potentials
although the
$\beta$ dependence is a little different.
 For the exponential potential, the set of allowed parameters is cut off abruptly in the 
 $\beta-\log_{10}U_0$ plane
 (see in the left-hand plots in fig. \ref{Exp_Better_ANk_NMax})
  when $\beta$ and $U_0$ are large as the Hubble scale (and by extension, $P_{\rm amp}$) becomes too large.
 In the equivalent plot for the quadratic potential (not shown) $H$ depends also on the displacement of the field from its true minimum, and so the effect of $U_0$ and 
$\beta$ is modulated by other parameters and the cutoff is not so clear. 

We have remarked that the differences between the observationally viable regions of the 
parameter spaces for the exponential and quadratic potentials are manifest
when the parameters $\beta$ and $\eta$ are plotted. It is interesting to compare the
plots in fig. \ref{Quad_Best_Ak} with those in fig. \ref{Best_ExpQuadCompare}, representing the
constraints derived analytically (in \cite{vandeBruck:2010yw}) under the assumption that the fields are dominated by their potential terms.
As shown in \protect\cite{vandeBruck:2010yw} the condition  $\rho_{\rm DBI}>\rho_{\varphi}$ 
is given by $W(x\approx 8\beta^2Ve^{4\beta\eta}/U_0)<2$ in terms of Lambert's W function for the
quadratic potential and $\eta>4\beta$ for the exponential potential.
It can be seen that the areas in fig. \ref{eta_beta_plots} that correspond to the shaded regions in fig. \ref{Best_ExpQuadCompare}
--- in which the DBI energy density is dominant ---
are almost completely devoid of points. Also, many of the points are clustered in the
vicinity of the $\rho_{\rm DBI}=\rho_{\varphi}$ boundary.
Thus, in the majority of runs with parameters that give rise to a power spectrum with a spectral index
within observable limits,
 the energy density of the DBI field is a subdominant but non-negligible fraction of the total energy density.

\bibliographystyle{JHEP}
\bibliography{DBI2Refs}

\end{document}